\documentclass[fleqn,usenatbib]{mnras}

\usepackage{newtxtext,newtxmath}

\usepackage[T1]{fontenc}

\DeclareRobustCommand{\VAN}[3]{#2}
\let\VANthebibliography\thebibliography
\def\thebibliography{\DeclareRobustCommand{\VAN}[3]{##3}\VANthebibliography}

\usepackage{graphicx}	
\usepackage{amsmath}	
\usepackage{textcomp}
\usepackage{booktabs}
\usepackage[dvipsnames]{xcolor}
\usepackage{tabularx}
\usepackage{longtable}
\usepackage{multirow}
\usepackage{pdflscape} 
\usepackage{csvsimple}	
\usepackage{soul} 

\title{IllustrisTNG and S2COSMOS: possible conflicts in the evolution of neutral gas and dust}

\author[Jenifer S. Millard et al.]{
Jenifer S. Millard,$^{1}$\thanks{E-mail: jenifer.millard@astro.cf.ac.uk}
Benedikt Diemer,$^{2}$
Stephen A. Eales,$^{1}$
Haley L. Gomez,$^{1}$ \newauthor
Rosemary Beeston, $^{1}$
\& Matthew W. L. Smith $^{1}$
\\
$^{1}$School of Physics and Astronomy, Cardiff University, The Parade, Cardiff, CF24 3AA, UK\\
$^{2}$Department of Astronomy, University of Maryland, College Park, MD 20742, USA\\
}

\date{Accepted 2020 October 8. Received 2020 September 28; in original form 2020 August 12.}

\pubyear{2020}

\begin{document}
\label{firstpage}
\pagerange{\pageref{firstpage}--\pageref{lastpage}}
\maketitle

\begin{abstract}
We investigate the evolution in galactic dust mass over cosmic time through i) empirically derived dust masses using stacked submillimetre fluxes at 850$\mu$m in the COSMOS field, and ii) dust masses derived using a robust post-processing method on the results from the cosmological hydrodynamical simulation IllustrisTNG. We effectively perform a ‘self-calibration’ of the dust mass absorption coefficient by forcing the model and observations to agree at low redshift and then compare the evolution shown by the observations with that predicted by the model. We create dust mass functions (DMFs) based on the IllustrisTNG simulations from 0 $< z <$ 0.5 and compare these with previously observed DMFs. We find a lack of evolution in the DMFs derived from the simulations, in conflict with the rapid evolution seen in empirically derived estimates of the low redshift DMF. Furthermore, we observe a strong evolution in the observed mean ratio of dust mass to stellar mass of galaxies over the redshift range 0 $< z <$ 5, whereas the corresponding dust masses from IllustrisTNG show relatively little evolution, even after splitting the sample into satellites and centrals. The large discrepancy between the strong observed evolution and the weak evolution predicted by IllustrisTNG plus post-processing may be explained by either strong cosmic evolution in the properties of the dust grains or limitations in the model. In the latter case, the limitation may be connected to previous claims that the neutral gas content of galaxies does not evolve fast enough in IllustrisTNG.
\end{abstract}



\section{Introduction}
Despite constituting around only 0.1 per cent of the baryonic content of a galaxy by mass (e.g. \citealt{Vlahakis2005}; \citealt{Dunne2011}; \citealt{Clemens2013}; \citealt{Beeston2018}; \citealt{Driver2018}), cosmic dust has a profound effect on our ability to study and understand the evolution of galaxies. Dust obscures and absorbs the optical and ultraviolet (UV) light from stars, and re-emits this energy at far-infrared (FIR) and sub-millimetre (sub-mm) wavelengths. Dust is thought to have absorbed around half of the starlight ever emitted in the universe (e.g. \citealt{Puget1996}; \citealt{Fixsen1998}; \citealt{Dole2006}). 

Cosmic dust is believed to originate from three main sources: the stellar winds of evolved low-mass stars, such as asymptotic giant branch (AGB) stars (e.g. \citealt{Salpeter1974}; \citealt{Valiante2009}), the violent deaths of massive stars (supernovae explosions, e.g. \citealt{Dunne2003SneDust}; \citealt{Morgan2003}; \citealt{Matsuura2011}; \citealt{Gomez2012}; \citealt{DeLooze2017, DeLooze2019}; \citealt{Chawner2019, Chawner2020}; \citealt{Cigan2019}), and grain-growth in the ISM (e.g. \citealt{Dwek2007}; \citealt{Draine2009}; \citealt{Dunne2011}; \citealt{Asano2013}; \citealt{Zhukovska2016}; \citealt{deVis2017a, deVis2017b}). Since dust is a by-product of star formation, studies examining cosmic dust allow us to probe the past star-formation history of a galaxy. Furthermore, since cosmic dust is an important constituent of the ISM and pervasive throughout it, dust has been found to be useful as a tracer to estimate the molecular gas content of galaxies, the fuel for future star formation. The optically thin dust continuum emission detected at a single sub-mm wavelength can be used as a tracer of molecular gas (\citealt{Eales2012}; \citealt{Scoville2014, Scoville2016, Scoville2017}; \citealt{Tacconi2018}; \citealt{Millard2020}). 

To enable studies of the gas and dust content of galaxies over much of cosmic history, astronomers have conducted sub-mm surveys of the sky. For example, the \textit{Herschel Space Observatory} (hereafter \textit{Herschel}, \citealt{Pilbratt2010}) was used in large-scale FIR surveys covering well-studied areas of sky (e.g. H-ATLAS, \citealt{Eales2010}), and the Submillimetre Common-User Bolometer Array 2 (SCUBA-2) on the James Clerk Maxwell Telescope (JCMT, \citealt{Holland2013}) has recently been used to study the COSMOS field \citep{Scoville2007} at unprecedented depths at sub-mm wavelengths \citep{Simpson2019}. The Atacama Large Millimetre Array (ALMA) has also been used to observe the dust continuum of a significant number of high-mass galaxies (e.g. \citealt{Scoville2017}; \citealt{Tacconi2018}, and references therein).

Despite several studies illustrating the usefulness of sub-mm dust emission as a tracer for gas mass (e.g. \citealt{Bethermin2015}; \citealt{Genzel2015}; \citealt{Scoville2016, Scoville2017}; \citealt{Tacconi2018}; \citealt{Millard2020}), such calculations involve significant assumptions. For example, the studies implicitly assume that metallicity does not evolve with redshift and that the fraction of metals incorporated in dust grains is a constant, or equivalently that the dust-to-gas ratio is constant with time. A more direct measurement to make with sub-mm data is to measure dust masses, in which the only assumption is that the properties of dust are constant in space and time.  

One of the fundamental measurements that can be made to describe the dust content of galaxies, particularly at different cosmic epochs, is the dust mass function (DMF) which is defined as the space density of galaxies as a function of dust mass \citep{Beeston2018}. The first measurements of the local DMF were made using sub-mm observations at 450$\mu$m and 850$\mu$m, using an indirect technique and were hampered by small number statistics (\citealt{Dunne2000}; \citealt{Dunne2001}; \citealt{Vlahakis2005}). Early efforts to probe beyond the local universe ($z=1$, \citealt{Eales2009}; $z=2.5$, \citealt{Dunne2003}) also suffered from poor statistics and a variety of other experimental limitations.

The advent of {\it Herschel}, with its better sensitivity and faster mapping speed, led to the first estimates of the DMF with good statistics. Using sub-mm data at 250$\mu$m on 1867 sources out to $z=0.5$, \citet{Dunne2011} found evidence for dramatic evolution in the dust content of galaxies, with five times more dust observed at $z=0.5$ than $z=0$ i.e. over the past 5 billion years of cosmic history. More recently, \cite{Beeston2018} presented an estimate of the local DMF with the best statistics to-date, using 15,750 galaxies in the equatorial GAMA fields \citep{Driver2011} out to a redshift of 0.1. \citet{Driver2018} later constructed an optically-selected sample to probe the dust density out to redshifts of $5$, finding little evolution from $0<z<0.5$, in contrast to \citet{Dunne2011}. \citet{Driver2018} found an increasing dust mass content at higher redshifts, with dust mass density peaking at $z=1$.

Although more detailed observations of the Universe over the past two decades have led to a deeper understanding of the content of galaxies, we still lack knowledge of the underlying physics governing galaxy evolution. This information is difficult to garner from observations alone. One solution to this problem is the development of suites of complex cosmological hydrodynamical simulations. Modern efforts include IllustrisTNG, which evolves a mock universe from shortly after the Big Bang through to the present day, including many of the physical processes that drive galaxy evolution, such as gas radiative mechanisms, star formation in the dense ISM, stellar population evolution, chemical enrichment, and feedback and outflows (\citealt{Marinacci2018}; \citealt{Naiman2018}; \citealt{Nelson2018}; \citealt{Pillepich2018a}; \citealt{Springel2018}). 

However, cosmological hydrodynamical simulations such as IllustrisTNG are tuned to reproduce a selected set of observational constraints, such as the stellar mass content of galaxies and cosmic star formation rate density at $z=0$ \citep{Pillepich2018a}. Therefore, it is vital to test these simulations by comparing their predictions to properties that they have not been tuned to reproduce.

Whilst current simulations, such as IllustrisTNG, do not model dust (although such models are in preparation e.g. \citealt{McKinnon2019}), dust properties can be examined using post-processing analysis (e.g. \citealt{Schulz2020}; \citealt{Vogelsberger2020}). A recent example of this is presented in \cite{Baes2020}, who examined the infrared luminosity functions and dust mass functions for galaxies from the EAGLE cosmological simulation (\citealt{Crain2015}; \citealt{Schaye2015}) for $z<1$. Synthetic multi-wavelength observations for EAGLE galaxies were generated using the radiative transfer code SKIRT (\citealt{Baes2011}; \citealt{CampsBaes2015}; \citealt{Camps2016, Camps2018}). Dust masses were then estimated by fitting a simple modified blackbody model to the synthetic luminosities generated using SKIRT at wavelengths of 160, 250, 350, and 500 $\mu$m. They found that EAGLE predicted only mild evolution in the DMF out to $z=1$, in contrast with the observations of the DMF of \cite{Dunne2011} but consistent with the milder evolution in the dust density found by \cite{Driver2018}. 

In our first study, (\citealt{Millard2020}, Paper I), we used the sub-mm emission from dust as a tracer of extra-galactic gas content over much of the history of the Universe. We estimated the average molecular gas mass fractions of galaxies in the COSMOS field using stacked 850$\mu$m fluxes and the gas scaling relations derived in \cite{Scoville2016, Scoville2017}. Stacking combines the fluxes from similar galaxies whose individual data signals are otherwise buried beneath noise to allow estimations of average galactic properties. Information on individual galaxies is lost, but is gained on the population as a whole. We calculated inverse-variance weighted average sub-mm fluxes for $\sim$63,000 binned sources with stellar masses 9.5 $\leq$ log$_{10}$($M_*/M_{\odot}$) $\leq$ 12, out to $z < 5$, extending to higher redshifts than previous studies, and including more `normal' star-forming galaxies on the Main Sequence (MS, e.g. \citealt{Daddi2007}; \citealt{Karim2011}; \citealt{Whitaker2012}; \citealt{Madau2014}; \citealt{Lee2015}). 

In this study (Paper II), we instead use the average stacked sub-mm fluxes from Paper I to examine the evolution of dust mass in galaxies over cosmic time out to $z<5$, which requires only the assumption that the properties of dust are a constant. We compare our observational results to the evolution of dust mass in galaxies predicted by IllustrisTNG using a post-processing analysis, to examine how well modern simulations trace dust. We also calculate the dust mass functions for IllustrisTNG galaxies, and compare these to the observations of \cite{Dunne2011} and \cite{Beeston2018}. 

We use the cosmological parameters from \textit{Planck} \citep{Planck2015} and make use of \texttt{astropy.cosmology} (\citealt{Astropy2013}; \citealt{Astropy2018}) assuming \texttt{FlatLambdaCDM} and $H_0 = 67.7 {\rm km \ Mpc^{-1} s^{-1}}$,  $\Omega_{m,0} = 0.307$, and $\Omega_{b,0} = 0.0486$.

\section{Data}
Here, we will briefly summarize the observational data used to derive average dust properties of physical galaxies over cosmic time, and also describe the simulations from IllustrisTNG used in this study. The observational datasets and catalogues used to compare with the simulations are described in more detail in \citet{Millard2020}.

\subsection{S2COSMOS map}
The JCMT is equipped with a 10,000 pixel bolometer camera which operates in the sub-millimetre (sub-mm) regime, specifically at wavelengths of 450$\mu$m and 850$\mu$m. This camera, SCUBA-2 (\citealt{Holland2013}), was recently used to produced the deepest and most sensitive 850$\mu$m map of the COSMOS field \citep{Scoville2007}, as part of the SCUBA-2 COSMOS (S2COSMOS; \citealt{Simpson2019}) survey. The map combines archival data from the SCUBA-2 Cosmology Legacy Survey (S2CLS, \citealt{Geach2017}; \citealt{Michalowski2017}), and new data from S2COSMOS, resulting in a complete and homogeneous map of the COSMOS field at 850$\mu$m. 

Instrumental noise dominates over confusion noise in the S2COSMOS survey; the median instrumental noise in the central 1.6deg$^2$ of the 850$\mu$m map is $\sigma_{850\mu {\rm m}}$ = 1.2mJy beam$^{-1}$. Additional coverage of 1deg$^2$ has a median instrumental noise level of $\sigma_{850\mu {\rm m}}$ = 1.7mJy beam$^{-1}$ \citep{Simpson2019}. In this work, we use the match-filtered map, which is more sensitive to point source emission. For further details of the maps produced in the S2COSMOS survey, we refer the reader to \cite{Simpson2019}. 

We derive the dust masses of galaxies in the S2COSMOS field by stacking on the  850$\mu$m map \citep{Millard2020}. As such, we require a source catalogue to provide the locations of COSMOS galaxies, which we briefly discuss in the next section.  

\subsection{Astrophysical source catalogues}
\label{Sec:2.2 sourcecats}
The COSMOS field is one of the most extensively studied areas of sky, with observations spanning almost the entire electromagnetic spectrum. As such, comprehensive photometric catalogues of the region have been created, and several spectral energy distribution (SED) fitting routines have been used with such data sets to constrain the physical properties of galaxies. In this work, we make use of the source catalogue presented in \cite{Driver2018}, wherein physical parameters of galaxies were estimated using the SED fitting routine {\sc magphys} \citep{daCunha2008}. We also use the COSMOS2015 catalogue from \cite{Laigle2016}. 

\subsubsection{{\sc magphys} catalogue}
{\sc magphys} (\citealt{daCunha2008}) produces probabilistic estimates of the physical properties of galaxies by using a $\chi^2$ minimization techique to fit pre-determined libraries of physically motivated model SEDs to panchromatic photometry data. Examples of the possible physical parameters returned include stellar mass estimates and star formation histories. The model SEDs of {\sc magphys} cover galactic emission from the UltraViolet (UV) through to the Far-Infared (FIR). Stellar emission is modelled using synthetic spectra from \citealt{BruzualCharlot2003}, assuming a \citealt{Chabrier2003} stellar initial mass function (IMF). This starlight can then be attenuated by dust present within the galaxy; either in spherically symmetric birth clouds, wherein lies a `warm' dust component with temperature 30-60\,K, or in the ambient interstellar medium (ISM), where the `cold' dust component is assumed to have a temperature of 15-25\,K \citep{CharlotFall2000}. {\sc magphys} employs an energy balance between the Ultraviolet-to-Near-Infrared (UV-NIR) and the Mid-Infrared-to-Far-Infrared (MIR-FIR) model components - the stellar emission attenuated by dust must be re-emitted in the FIR. 

In this work, we use the {\sc magphys} catalogue ({\tt MagPhysG10v05}) of \citealt{Driver2018}, which is based on the 22-band panchromatic photometry catalogue of \citealt{Andrews2017} and contains 173,399 sources. The photometric catalogue used with {\sc magphys} ({\tt G10CosmosLAMBDARCatv06})  is $i$-band < 25 mag limited \citep{Driver2018}. As stated in \cite{Andrews2017}, the source catalogue is complete for objects with $i$-band < 24.5 mag and partially complete to $i$ < 25 mag. 

Redshifts for the {\sc magphys} catalogue are sourced from \cite{Davies2015}, who performed an independent extraction of spectroscopic redshifts from the zCOSMOS-Bright sample \citep{Lilly2007zbright} and combined these redshifts with archival spectroscopic redshifts from the PRIMUS, VVDS and SDSS (\citealt{Cool2013}; \citealt{LeFevre2013}; \citealt{Ahn2014}) surveys. If spectroscopic redshifts were not available, sources were assigned photometric redshifts from COSMOS2015 \citep{Laigle2016}.

\subsubsection{COSMOS2015 catalogue}
The COSMOS2015 catalogue from \cite{Laigle2016} contains half a million NIR selected objects with photometry from the X-ray range, through to the radio. In the deepest $K_s$-band regions, for stellar masses >10$^{10} M_{\odot}$, the catalogue is 90\% complete out to $z$=4. Where possible, sources in the catalogue have spectroscopic redshifts; otherwise photometric redshifts ($\sigma_{\Delta z / (1+z)}$ = 0.007 for $z < 3$, and $\sigma_{\Delta z / (1+z)}$ = 0.021 for 3 < $z$ < 6) are assumed. 

\subsubsection{Final astrophysical source catalogue}
\cite{Andrews2017} advise that sources in their catalogue without matches in COSMOS2015 \citep{Laigle2016} should be treated with caution. We cross-match the {\tt MagPhysG10v05} catalogue \citep{Driver2018} to the publicly available COSMOS2015 catalogue \citep{Laigle2016} on RA and Dec, to ensure catalogue completeness, and to extend the wavelengths for which photometric measurements are available. The cross-matched {\sc magphys}-COSMOS2015 catalogue covers the central 1 deg$^2$ of the S2COSMOS map, where instrumental noise is lowest.

Emission from Active Galactic Nuclei (AGN) can contaminate measurements of galactic emission, particularly in the infrared, and subsequently systematically impact estimates of physical parameters like stellar masses and star-formation rates \citep{Ciesla2015}. Since disentangling galactic emission from that originating from AGN is non-trivial, we choose to remove AGN from the cross-matched {\sc magphys}-COSMOS2015 catalogue. We identify AGN using IR, radio and X-ray data, and also remove sources with {\sc magphys} stellar masses greater than $10^{12}M_{\odot}$. A full description of this process is available in \cite{Millard2020}.

In this work, we limit ourselves to considering galaxies in the {\sc magphys}-COSMOS2015 catalogue with log$_{10}(M_*/M_{\odot}) \geq 9.5$. In total, our final {\sc magphys}-COSMOS2015 astrophysical source catalogue, without AGN, consists of 63,658 galaxies, of which 13,955 have reliable spectroscopic redshifts\footnote{For consistency the redshifts used in this study are from \cite{Driver2018}, since these are the redshifts used in producing the {\sc magphys} estimates of the stellar masses.}.  We note that the {\sc magphys}-COSMOS catalogue provides the RA, Dec, $M_*$ and $z$ for our observed galaxies. We do not use the {\sc magphys} dust masses in this work.

\subsection{IllustrisTNG} \label{Sec:2.3 IllustrisTNG}
IllustrisTNG is a suite of state-of-the-art hydrodynamical cosmological simulations designed to illuminate the underlying physical processes that drive galaxy formation (\citealt{Marinacci2018}; \citealt{Naiman2018}; \citealt{Nelson2018}; \citealt{Pillepich2018a}; \citealt{Springel2018}). The simulations model both baryons and dark matter, and encode many physical processes, such as stellar evolution, galactic winds, and chemical enrichment schemes (\citealt{Weinberger2017}; \citealt{Pillepich2018a}; \citealt{Pillepich2018b}), to try and create realistic representations of the universe from the Big Bang until the present day. IllustrisTNG utilises the moving mesh code {\sc arepo} \citep{Springel2010} to model the evolution of galaxies over cosmic time. Galaxies and halos are identified using the {\sc subfind} algorithm (\citealt{Davis1985}; \citealt{Springel2001}; \citealt{Dolag2009}). The updated galaxy formation model used in IllustrisTNG is based on the one originally presented in the Illustris simulation (\citealt{Vogelsberger2013, Vogelsberger2014b, Vogelsberger2014a}; \citealt{Genel2014}; \citealt{Torrey2014}; \citealt{Sijacki2015}). The Illustris TNG simulations follow the cosmology of the \cite{Planck2015}.

The simulations are run under three different resolutions - in this work, we focus on TNG100, with a box side length of 106.5Mpc. We consider snapshots from the present day, out to $z$=5, such that we explore a similar redshift range compared to our observational data. Of particular interest to this study are the predictions for the evolution of the neutral gas and metallicity (e.g. \citealt{Nelson2018b}; \citealt{Pillepich2018a}; \citealt{Diemer2018}; \citealt{Diemer2019}).

\subsubsection{Simulated galaxies from TNG100}\label{sec:2.3.1_galaxiestng100}
In TNG100, there are a few objects that may be artifacts (\citealt{Nelson2019}). From our sample, we therefore remove galaxies with zero neutral gas and no metals (these are mostly satellite galaxies that have presumably undergone excessive stripping through galaxy-galaxy interactions; see \citealt{Diemer2019}).

When comparing to data from the literature, we impose a lower stellar mass limit of 200 times the baryonic mass resolution of the simulation (200$m_b$; $m_b = 1.4 \times 10^6 M_{\odot}$) to avoid galaxies that may be poorly resolved in the simulations and therefore not reliably modelled (\citealt{Shen2020}). In other words, we perform a stellar mass selection on IllustrisTNG galaxies, without considering other galactic baryonic mass components e.g. gas mass. When comparing to only our stacking data, we consider only TNG100 galaxies with stellar masses within the bounds imposed by the astrophysical data i.e. $9.5 \leq {\rm log}(M_*/M_{\odot}) \leq 12$.

In this study, we use parameter values for a galaxy that are calculated for particles and cells within two stellar half-mass radii ($2r_{0.5}$). However, to ensure the robustness of our conclusions and to check that our choice of TNG100 dataset does not influence our conclusions, we also perform the same calculations (re-running all results) but for parameter values calculated using all particles gravitationally bound to the galaxy. We do not discuss these results in the paper, other than to say we find that using parameter values calculated for all particles and cells gravitationally bound to a source does not change our conclusions. Table \ref{tab:TNGnumbers} illustrates the number of TNG100 galaxies considered at each redshift, for the $2r_{0.5}$ dataset.

\begin{table}
    \centering
    \caption{Number of TNG100 galaxies considered at each redshift, after removal of un-physical galaxies and resolution mass cuts to 200$m_b$ (N$_{\rm filt}$), and after additional mass-cuts to allow for comparison with our observations obtained from stacking galaxies in the S2COSMOS field (N$_{\rm filt, obs}$) (see Sections \ref{Sec:2.2 sourcecats} and \ref{Sec:2.3 IllustrisTNG}).} 
    \begin{tabular}{cccc}
    \hline
    \multirow{2}{*}{Snapshot} & \multirow{2}{*}{$z$} & \multicolumn{2}{c}{$2r_{0.5}$} \\
    \cline{3-4}
     & & $N_{\rm filt}$ & $N_{\rm filt, obs}$ \\
    \hline
    17 & 5 & 3185 & 300 \\ 
    21 & 4 & 7394 & 996 \\ 
    25 & 3 & 14338 & 2631 \\ 
    33 & 2 & 22902 & 5531 \\ 
    40 & 1.5 & 26270 & 7227 \\ 
    50 & 1 & 27456 & 8680 \\ 
    59 & 0.7 & 27442 & 9350 \\ 
    72 & 0.4 & 26734 & 9563 \\ 
    78 & 0.3 & 26393 & 9513 \\ 
    84 & 0.2 & 26168 & 9462 \\ 
    91 & 0.1 & 25660 & 9298 \\ 
    99 & 0 & 25254 & 9252 \\ 
    \hline
    \end{tabular}
    \label{tab:TNGnumbers}
\end{table}

\subsection{Stellar mass distributions}\label{sec:2.4_stellmasshist}
Before we compare dust masses for the observational and simulated data, we perform a quick check and examine the normalised stellar mass distributions of galaxies in the two datasets. We bin the galaxies in the observational {\sc magphys}-COSMOS catalogue into the same redshift bins as the IllustrisTNG snapshots (Figure \ref{fig:mstar_hists_stacking_TNG} and Table \ref{tab:TNGnumbers}). We see that broadly, the stellar mass distributions are similar. We can be confident that the observed and simulated galaxies represent similar galactic populations.

\section{Dust masses}
\subsection{Empirical dust masses from stacking on S2COSMOS 850\texorpdfstring{$\mu$}{u}m map}
Typically, the 850$\mu$m emission from individual galaxies is below the noise level of the S2COSMOS map \citep{Simpson2019}. Therefore, in Paper I, we studied the 850$\mu$m emission of galaxies in the COSMOS field in a statistical manner - we binned galaxies by redshift and stellar mass to create sub-populations of these sources, and used well-established inverse-variance weighted (IVW) stacking methodologies to estimate mean sub-mm fluxes for sources in each bin. Although this does mean that we lose information on individual galaxies, we gain information on each sub-population as a whole. We estimated 850$\mu$m stacked flux errors using Monte Carlo (MC) simulations on the location of galaxies within the S2COSMOS map. See \cite{Millard2020} for full details. 

For sub-mm emission measured on the Rayleigh-Jeans tail of dust emission, dust masses, $M_d$, can be estimated using:
\begin{equation} \label{Eq:Md_stack}
    M_d = \frac{<S_{\nu_o}> D_L^2}{\kappa_{\lambda_e} B_{\nu_e}(T_d) (1+z)}
\end{equation}
where $<S_{\nu_o}>$ is the average stacked flux for galaxies in a given $(M_* - z)$ bin, measured at the wavelength of observation (in this case, 850$\mu$m), $D_L$ is the luminosity distance of the source\footnote{In this work, we make use of the {\tt Python} package {\tt astropy.cosmology}, assuming {\tt Planck15} cosmology, to calculate the luminosity distance using the centre of the redshift bin for the sub-population in question.}, $\kappa_{\lambda_e}$ is the assumed dust mass absorption coefficient scaled to the rest-frame emission wavelength, $B_{\nu_e}(T_d)$ is the Planck function calculated at the rest-frame emission frequency for an assumed dust temperature, $T_d$, and $z$ is the middle of the redshift bin under consideration (e.g. \citealt{Dunne2003}). 

The value $\kappa_{\lambda_e}$ is notoriously uncertain. Indeed, a recent study by \cite{Clark2016} showed that estimates of $\kappa_{500}$ in the literature span over 3.5 orders of magnitude. We have used the freedom produced by this uncertainty to choose a value of $\kappa_{500}$ that gives as good agreement as possible between the observed low-redshift DMFs (from \citealt{Dunne2011} and \citealt{Beeston2018}) and the one predicted by IllustrisTNG (Section \ref{sec:4.1_localDMF}). Rather than a formal fit, given the uncertainties and small differences in the shape of the predicted and observed $z=0$ DMFs, we determined $\kappa_{500}$ by performing a rough fit by eye to the high-mass end of the $z=0$ DMFs. In particular, we ensured a good fit with the $z<0.1$ DMF from \cite{Dunne2011}. By performing this `self-calibration' of $\kappa_{500}$ at low redshift, we are still able to test whether IllustrisTNG predicts the correct shape of the low-redshift DMF and whether it predicts the cosmic evolution seen in the observations. We assume $\kappa_{500} = 0.14$ m$^2$kg$^{-1}$, which is close to that found by \cite{James2002} and roughly around the middle of the range of values given by \cite{Clark2016}. Note that we scale any dust masses taken from the literature to our chosen $\kappa_{500}$ values. We scale the dust mass emissivity coefficient to the rest frame emission wavelength using:
\begin{equation} \label{Eq:scale_kappa}
    \kappa_{\lambda_e} = \kappa_{500} \left( \frac{\lambda_{500}}{\lambda_e} \right)^\beta
\end{equation}
where $\beta$ is the dust emissivity spectral index. We assume $\beta$ = 1.8 \citep{Planck2015}, consistent with our previous stacking work and other studies upon which this work was based (see \citealt{Millard2020} for details; see also \citealt{Scoville2016}; \citealt{Scoville2017}). Once again, for our stacked data from the S2COSMOS field, the rest frame emission wavelength is estimated by using the middle of the redshift bin currently being investigated. By scaling the dust mass emissivity coefficient to the wavelength of emission in the rest frame, we implicitly assume that the properties of dust grains themselves are constant throughout the universe. We assume a mass-weighted dust temperature of 25\,K, as in \cite{Millard2020} (see also \citealt{Scoville2016}; \citealt{Scoville2017}). We note that our method for calculating dust masses is simple, only
depends on a few parameters, and does not have the black-box complexity of spectral energy distribution-fitting methods such as {\sc magphys}. We discuss the impact of our assumptions on the values for $\beta$, $\kappa_{500}$ and mass-weighted dust temperature in Section \ref{sec:5.1_caveats_assumps}.

\subsection{Dust masses from IllustrisTNG} \label{sec:3.2_dustmasses_TNG_eqn}

Dust masses are not one of the physical parameters explicitly calculated for galaxies in IllustrisTNG. Therefore, we have to calculate them in post-processing. We calculate dust masses, $M_{d,{\rm TNG}}$, on a cell-by-cell basis, summing over all cells within two stellar half-mass radii of a given galaxy:
\begin{equation} \label{Eq:TNG-MD}
    M_{d,{\rm TNG}} = \varepsilon_d \left( \sum_{i=1}^{N_{2r}} \frac{f_{H, {\rm neutral},i}}{f_{H,i}} \, M_{{\rm gas},i} \, Z_i \right)
\end{equation}
where $M_{{\rm gas},i}$ is the mass of gas in a given cell, $Z_i$ is the metallicity of the gas in the cell, $f_{H,\rm neutral,i}$ is the fraction of gas in the cell that is neutral hydrogen (atomic or molecular hydrogen; HI or H$_2$), $f_{H,i}$ is the fraction of all the gas in a cell that is hydrogen and $N_{2r}$ is the number of cells within two stellar half-mass radii for a given source. $\epsilon_d$ is the fraction of metals in a galaxy's ISM that are locked up in dust, a free parameter. We justify our choice of $\varepsilon_d$ at the end of this Section.

In IllustrisTNG, the neutral gas fraction is calculated using two different prescriptions, depending on whether cells are classified as star-forming or not. For cells below the star formation threshold density, the neutral gas fraction is calculated self-consistently within the simulations (\citealt{Vogelsberger2013}). The combination of the cooling rate (from primordial hydrogen and helium cooling, metal-line cooling and inverse Compton cooling off CMB photons), the photoionisation rate (from the ultraviolet background; UVB) and gas self-shielding provide an overall UVB photoionization rate, which is used with {\sc cloudy} look-up tables (\citealt{Ferland1998}) to calculate the neutral gas fraction. However, for star-forming cells where the gas density is higher, these approximations break down; the neutral gas fraction is not calculated self-consistently within the simulations (\citealt{SpringelHernquist2003}). For these cells, the model splits gas into a hot and cold phase, where the cold phase is assumed to be entirely neutral gas. The fraction of neutral hydrogen in star-forming cells is then estimated based on the density fraction of cold gas clouds, with a typical fraction between 0.9 and 1 (\citealt{SpringelHernquist2003}).

\subsubsection{A constant dust-to-metal ratio?}
\label{sec:3.2.1_constant_epsilon_d}

\begin{figure}
	\includegraphics[width=\columnwidth]{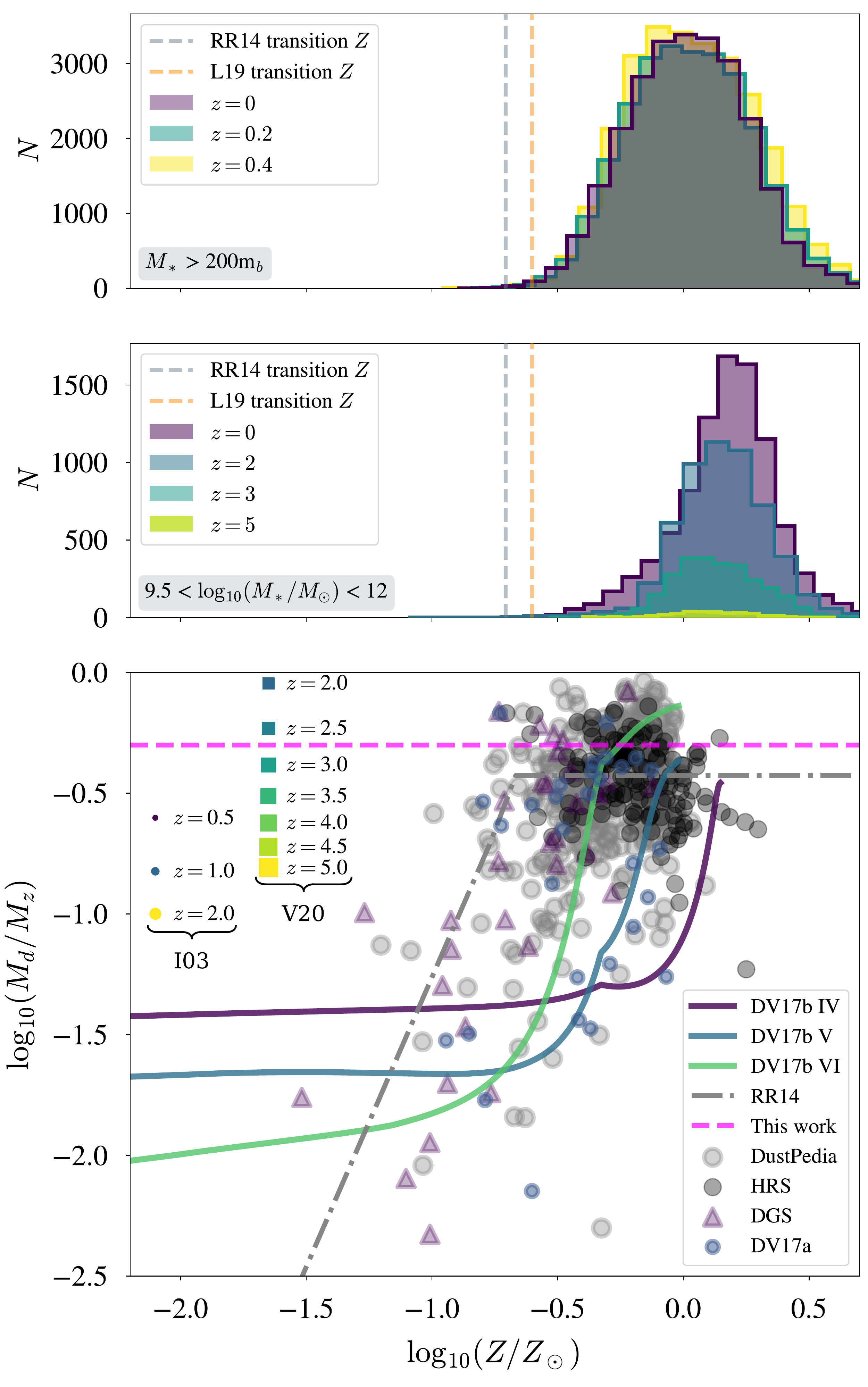}
       \caption{The dust-to-metal ratio for the local Universe and our samples. In the top two panels, we show the distribution of the metallicities of IllustrisTNG galaxies for the two sub-samples we use in this work. The vertical lines indicate the transition metallicity from \citet{Remy-Ruyer2014} {\it (grey)} and from \citet{Lagos2019} {\it (orange)}.
       In the bottom panel, we show local galaxy samples DustPedia \citep{devis2019}, the gas-selected HSS sample \citep{deVis2017a,deVis2017b}, the revised parameters for the Dwarf Galaxy Sample (DGS, \citealt{Madden2013}) from \citet{deVis2017b} and  the {\it Herschel} Reference Survey (HRS, \citealt{Boselli2010}).  Following \citet{devis2019}, the metal mass $M_Z$ is defined as
$M_Z = (Z \times M_{\rm gas}) + M_{\rm dust}$ where the fraction of metals in the ISM by mass $Z$ is defined as $Z = 27.67 \times 10^{12+{\rm log(O/H)}-12}$ where the solar value of $12+\rm log(O/H)$ is $8.69$.  Following IllustrisTNG, we use a Solar value of $Z_{\odot} = 0.0127$ \citep{Wiersma2009}.  A broken power-law relationship with $Z$ from \citet{Remy-Ruyer2014} (their Table~1, using the Milky-Way $\rm CO-H_2$ conversion factor) is shown as the {\it dot-dashed line}. Predictions of the evolution of dust-to-metal ratios are illustrated by the models from \citet{deVis2017b} ({\it solid lines}). {\it Coloured circles} show the range of the dust-to-metals predicted as a function of $z$ from \citet{Inoue2003} (I03), with markers increasing in size and colour-coded by redshift. {\it Square coloured boxes} show the values proposed to match observed UV luminosity functions from $2<z<9$ \citep{Vogelsberger2020} (V20). These are included to show potential variations in $\epsilon_d$ with redshift. }
    \label{fig:dust_metals}
\end{figure} 

The most direct estimates for the dust-to-metal ratio, $\epsilon_d$, come from using UV spectroscopic observations of stars to measure the metal abundances within stars and in the ISM, the difference between the two being attributed to the metals depleted from the interstellar gas and locked up in dust grains (\citealt{James2002}). \citet{Clark2016} list 12 different estimates of $\epsilon_d$, which have a mean value of 0.5 and a standard deviation of 0.1. Most of these measurements, however, are ultimately based on depletion measurements in the Milky Way or rely on theoretical models of dust grains. The only other galaxies for which depletion measurements have been made using UV spectroscopy of stars are the Magellanic Clouds (\citealt{RomanDuval2019}), which have much lower stellar masses than the galaxies in our COSMOS sample. The depletion measurements imply that $\epsilon_d$ in these galaxies is lower than in the Milky Way, by a factor of 1.5 for the LMC and a factor of 4 for the SMC.  In an alternative approach to this question, \citet{Remy-Ruyer2014} and \citet{deVis2017b,devis2019}, in studies of large samples of nearby galaxies, find a linear relationship between dust-to-gas ratio and metallicity
for galaxies with $12+{\rm log(O/H)} > 8.0$.
The former work implies a constant value of $\epsilon_d$ of $\sim$0.3 and the latter implies $\sim$0.2 for evolved galaxies\footnote{We note that these quoted values for $\epsilon_d$ were calculated assuming a different dust mass emissivity coefficient compared to this study. 
Scaling to the same $\kappa_{\lambda}$ as used in this work results in a dust-to-metal ratio $\sim$20 and $\sim$30 percent higher than quoted in the original studies, thus reducing any perceived difference between the absolute values of quoted $\epsilon_d$ here and in previous works.}, suggesting that only the lowest mass, lowest metallicity galaxies would likely deviate from this constant value locally. Both studies observed a broken power law where the dust-to-metals ratio starts to vary with metallicity below a `transition' value of $12+{\rm log(O/H)} \sim 8$. An illustrative comparison of the dust-to-metals ratio observed in the local Universe is shown in Figure~\ref{fig:dust_metals}, including predictions from the chemical evolution models of \citet{deVis2017b}. 

It is not immediately clear what values of $\epsilon_d$ are appropriate at high redshifts, where high stellar mass galaxies are at an earlier stage of their evolution and therefore may have lower metallicities. However, \citet{devis2019} found galaxies in the local DustPedia survey with high dust-to-metal ratios ($0.2-0.5$) even at early evolutionary stages (defined as galaxies with high gas fractions).  For the highest gas fractions in their sample ($f_{\rm gas} > 0.8$), they began to see dust-to-metal ratios as low as 0.04.  Observations at high redshifts include gamma-ray bursts and damped Lyman alpha systems (\citealt{DeCia2013}; \citealt{Zafar2013}) where the observed values of $\epsilon_d$ for high-metallicity galaxies is similar to that for
the Milky Way (see also \citealt{Yajima2014}).  \citet{DeCia2013} suggest that $\epsilon_d$ could be lower for very low-metallicity galaxies. Similarly, no obvious trend with redshift was seen by \citet{Wiseman2017}, but extreme changes in $\epsilon_d$ are postulated for galaxies at $z>6$ (see \citealt{Vogelsberger2020} for more discussion). \citet{Li2019} however see little evolution in the dust-to-metal ratio from $z=0$ to $z=6$, instead showing it is most sensitive to the gas phase metallicity $Z$ (though decreasing values are seen at lower stellar masses and high gas fractions; \citealp{deVis2017b}). 

For the redshift range used in this work, both the hydro-simulations from \citet{McKinnon2016} and semi-analytical models of \citet{Popping2019} imply that $\epsilon_d$ is different at $z=0$ and $z=2$.  More recently, using the IllustrisTNG simulations we use in this work, \citet{Vogelsberger2020} required a dust-to-metal ratio (their $f_Z$ parameter) to vary with redshift over the range $2< z <8$ in order to match observed UV luminosity functions. Their relationship ($\epsilon_d = 0.9 \times (z/2)^{-1.92}$) predicts $\epsilon_d$ values of 0.9 - 0.1 for the redshift range which overlaps with this work ($2<z<5$), see Figure~\ref{fig:dust_metals}. We note that this $z=2$ value is higher than that predicted by \citet{Inoue2003} (Figure~\ref{fig:dust_metals}) and \cite{McKinnon2016}.

Figure~\ref{fig:dust_metals} compares the distribution of metallicities in the two sub-samples of the IllustrisTNG simulations used in this work: the $200m_b$ cut used for comparing with local galaxies (redshifts $0<z<0.4$), and the COSMOS stellar-mass cut ($9.5 \leq {\rm log}(M_* /M_{\odot}) \leq 12$) we use for comparison with our stacking data out to redshift $z=5$. To investigate the the impact of metallicity on the dust estimates used in this work, we compare with the broken power-law of \citet{Remy-Ruyer2014} which suggest departures from a constant $\epsilon_d$ only occurs at $Z/Z_{\odot} < 0.2$ ($Z/Z_{\odot} < 0.25$ for \citealt{Lagos2019}). For the $200m_b$ low $z$ sample, some galaxies have metallicities equal to, or below, the \citet{Remy-Ruyer2014} and \citet{Lagos2019} transition metallicity, but these are small numbers - $N\sim 12$ ($N \sim 60$) for \citet{Remy-Ruyer2014} (\citealt{Lagos2019}). The majority have metallicities where a constant dust-to-metal ratio is appropriate.  For the COSMOS sample, very few of our sources are at low enough metallicities, even at high redshifts. In summary, we show there is observational evidence that both our local samples of IllustrisTNG galaxies and high-metallicity systems have a fairly constant value of $\epsilon_d$ over the redshift range appropriate in this work, with a value similar to that of the the Milky Way. 
Figure \ref{fig:dust_metals} illustrates that a constant value of $\epsilon_d$=0.5, the average value from the meta-analysis of \cite{Clark2016}, is appropriate for the TNG100 sources used in this study. We will return to this issue in Section~\ref{sec:5_discussion}.

Finally, we note that a similar method was used to derive dust masses for IllustrisTNG galaxies by \cite{Hayward2020}. They define ISM gas using a temperature-density cut from \cite{Torrey2012} (summing over all particles within 25kpc of the subhalo centre) and use this to compute an ISM metal mass, which they convert to a dust mass using a constant dust-to-metal ratio of 0.4.

\section{Results}

\subsection{Local DMF: IllustrisTNG} \label{sec:4.1_localDMF}
Before examining the evolution of dust mass in IllustrisTNG galaxies over much of cosmic time, we first investigate the dust content of IllustrisTNG galaxies in the local universe by calculating the IllustrisTNG DMF for $z < 0.5$. We compare these low redshift DMFs to ones from the literature - specifically, from \cite{Dunne2011} and \cite{Beeston2018}. \cite{Dunne2011} used a sample of 1867 sources selected at 250$\mu$m from the Science Demonstration Phase of H-ATLAS \citep{Eales2010} with reliable Sloan Digital Sky Survey (SDSS (DR7); \citealt{Abazajian2009}) counterparts. Dust masses are estimated by using both a single and multiple temperature modified blackbody models for the SEDs, assuming a $\kappa_d$ value consistent with that used in {\sc magphys}. Their sample covers a redshift range of $0<z<0.5$.
\cite{Beeston2018} uses {\sc magphys} to estimate dust and stellar masses for optically selected galaxies from the three equatorial fields of the Galaxy and Mass Assembly Catalogue (GAMA; \citealt{Driver2011}), which are the fields that also have FIR/sub-mm data from {\it Herschel}-ATLAS \citep{Eales2010}. In total, they consider 15750 sources covering around 145 deg$^2$ of the sky, at redshifts $z < 0.1$.

\begin{figure*}
	\includegraphics[width=0.9\textwidth]{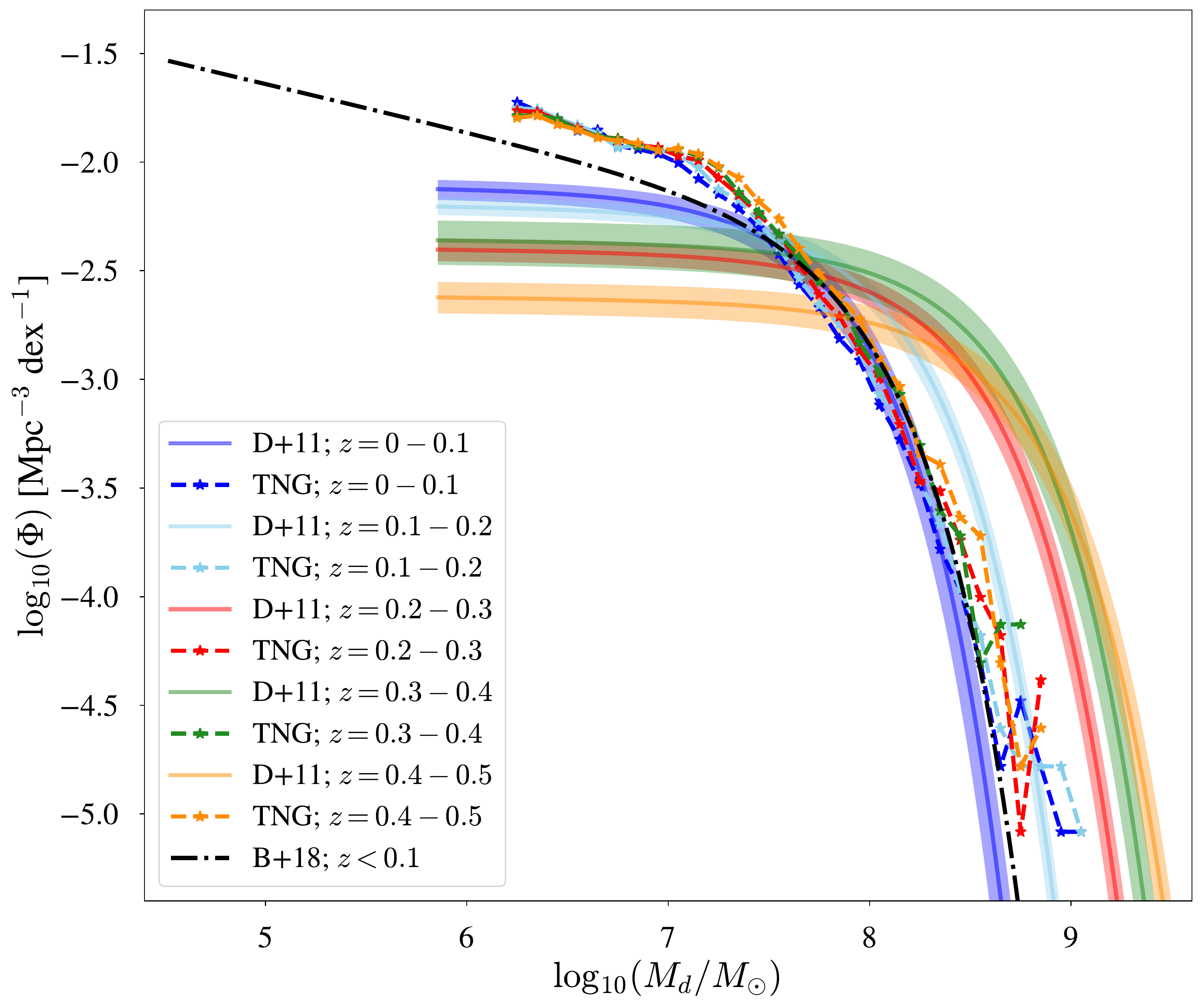}
    \caption{Low redshift DMFs for IllustrisTNG galaxies \textit{(stars and dashed lines)} and local DMFs from \protect\cite{Beeston2018} {\it (black dash-dot line)} and \protect\cite{Dunne2011} {\it (solid colour lines)}. The DMF from \protect\cite{Beeston2018} is based on an optically selected sample of 15750 galaxies within the redshift range $0.002 \leq z \leq 0.1$. The DMFs from \protect\cite{Dunne2011} are based on a 250$\mu$m selected sample of 1867 galaxies with redshifts $z < 0.5$. The width of the dot-dashed line represents the error for the \protect\cite{Beeston2018} DMF; the coloured shaded regions show the errors for the \protect\cite{Dunne2011} DMFs. The colours of the IllustrisTNG DMFs and DMFs from \protect\cite{Dunne2011} are chosen to indicate the same redshift bins. Local observed DMFs are scaled to IllustrisTNG cosmology with the dust masses scaled to $\kappa_{500}$.}
    \label{fig:lowz_DMF}
\end{figure*}

We calculate the IllustrisTNG DMFs over a similar dust mass range to those in the literature\footnote{Specifically, our lower dust mass limit in Figure \ref{fig:lowz_DMF} is dictated by the average dust mass of galaxies in the lowest stellar mass bin of Figure \ref{fig:logMd_logMs_z0}, to avoid using dust mass bins which are affected by the sharp stellar mass cut off of 200$m_b$ introduced in Section \ref{sec:2.3.1_galaxiestng100}.}, and bin our galaxies in stellar mass bins of 0.1 dex. We scale the DMFs from \cite{Dunne2011} and \cite{Beeston2018} to the same cosmology as assumed in the IllustrisTNG simulations, and to $\kappa_{500}$ (Figure \ref{fig:lowz_DMF}). Scaling to the $\kappa_{500}$ used in this study increases the dust masses quoted in the literature studies by a factor of 1.43. 

In Figure \ref{fig:lowz_DMF}, we see that at high dust masses, beyond the knee of the DMFs, the $z=0$ TNG DMF agrees well with the DMF from \cite{Beeston2018}, and with the $z<0.1$ DMF from \cite{Dunne2011}. We note that some of this agreement is by construction since we calibrated $\kappa_{500}$ by insisting on good agreement between the observed and theoretical DMF at high dust masses, but the agreement between the shapes of the DMFs (beyond the knee) is still significant. At lower dust masses, we find a higher number density of galaxies for TNG, compared to \cite{Beeston2018}. We attribute the excess number of sources at low dust masses to galaxies classified as `satellites' in the simulation, which have likely undergone excessive stripping (\citealt{Diemer2019}; \citealt{Stevens2019}), therefore reducing their dust masses as calculated using our post-processing technique and causing an excess of sources with low dust masses as compared to observations. We return to this issue and explore the differences between `satellite' and `central' TNG100 galaxies in Section \ref{sec:5.2_sats_cents}. However, the slope of the DMF at low dust masses seems similar to that found by \cite{Beeston2018}, even if there is an offset in the absolute number density. 

The most striking difference between the observational and theoretical DMFs (Figure \ref{fig:lowz_DMF}) is that we do not see the evolution in the DMF over the redshift range $0 < z < 0.5$ that is seen in the observational data. The TNG DMFs show little evolution up to $z = 0.5$, which is in stark contrast to the dust mass evolution shown in the observational data from \cite{Dunne2011}. However, it is worth noting that \cite{Dunne2011} caution that their DMFs at $z>0.3$ are estimated from galaxies that mostly only have photometric redshifts. Even so, the strongest evolution in the DMFs from \cite{Dunne2011} is at redshifts below this threshold, and this evolution is clearly not apparent in the TNG100 DMFs.

We now examine the $z=0$ ($M_* - M_d$) distribution of TNG100 galaxies and compare this distribution to local observational studies, as a sanity check that the dust mass estimates for TNG100 galaxies are consistent with those observed in the local galaxy population (Section \ref{sec:4.2_local_dust_mass}).

\subsection{Dust Mass Estimates at \texorpdfstring{$z$}{z} = 0} \label{sec:4.2_local_dust_mass}
\begin{figure*}
	\includegraphics[width=\textwidth]{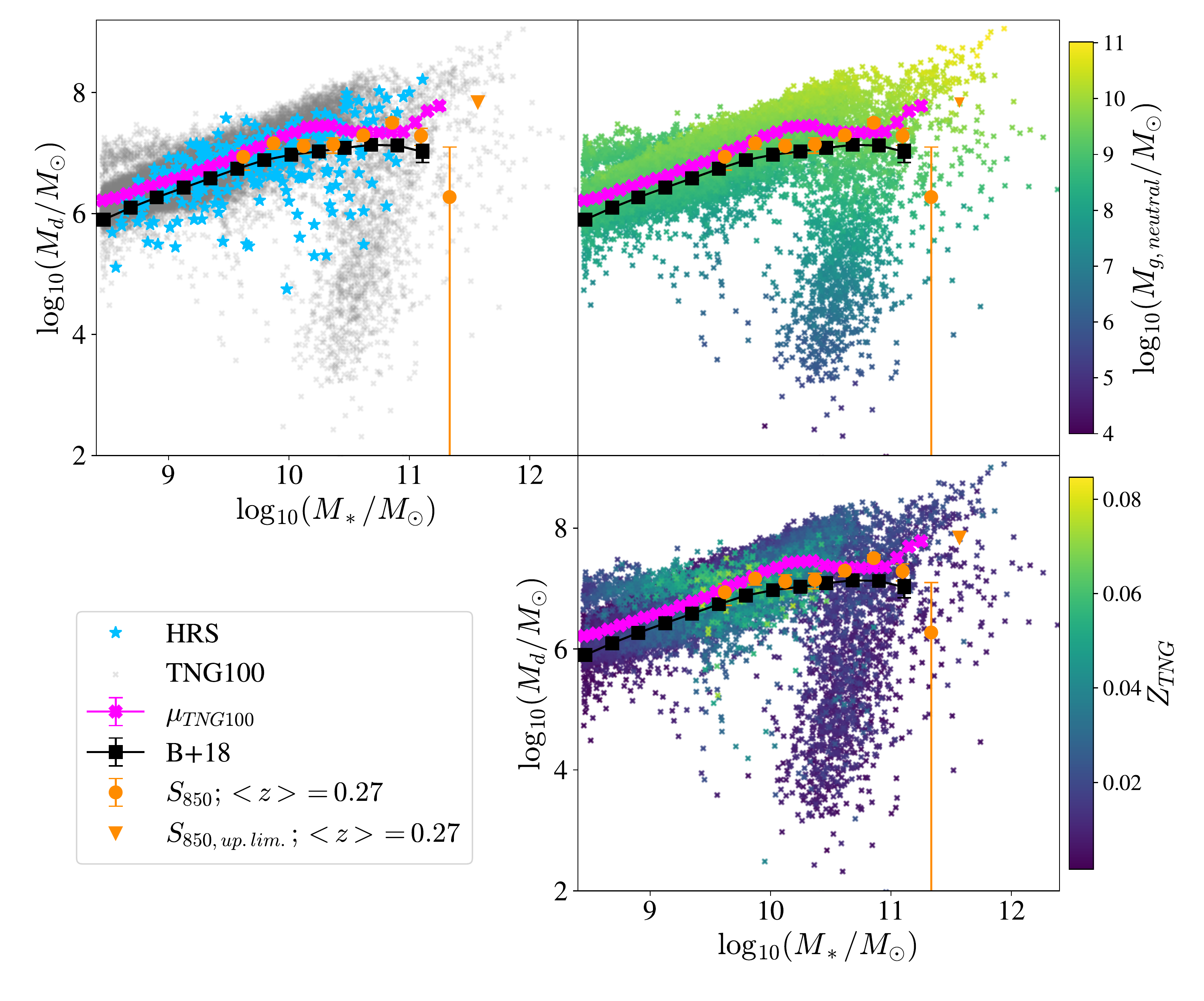}
    \caption{Dust mass versus stellar mass for the TNG100 galaxies at $z=0$ and
    for various low-redshift galaxy samples. The left-hand panel shows the TNG100 galaxies as {\it grey points}; the right-hand top panel shows them as crosses with a {\it colour} given by each galaxy's neutral gas mass (see colour bar); and the right-hand bottom panel shows them as crosses with a colour given by each galaxy's metallicity (see colour bar).
    The mean and standard error on the mean for the TNG100 galaxies in a given stellar mass bin are shown by {\it fuchsia crosses}. 
    The galaxies in the Herschel Reference Survey are shown as {\it blue stars} (\protect\citealt{deVis2017a}), the mean dust masses from the sample of \protect\cite{Beeston2018} as {\it black squares}, and the mean dust masses for the galaxies in the low-redshift bins from our S2COSMOS study as {\it orange circles}, with the {\it orange triangle} showing an upper limit.}
    \label{fig:logMd_logMs_z0}
\end{figure*}

We compare the dust masses calculated using the $z$=0 TNG100 data to dust masses calculated using our stacked S2COSMOS fluxes, and to dust masses from other studies of the local universe from \cite{deVis2017a} and \cite{Beeston2018} (Figure \ref{fig:logMd_logMs_z0}). \cite{deVis2017a} use {\sc magphys} to estimate dust and stellar masses for 323 galaxies that comprise the Herschel Reference Survey (HRS; \citealt{Boselli2010}). The HRS is a stellar-mass selected volume-limited sample of galaxies in the nearby Universe. Sources have distances between 15 and 25Mpc, and are $K$-band selected, to minimise dust selection effects caused by extinction. The sample contains both late- and early-type galaxies. A brief description of the data used in \cite{Beeston2018} can be found in Section \ref{sec:4.1_localDMF}. The dust masses from \cite{deVis2017a} and \cite{Beeston2018} are scaled to $\kappa_{500}$. In Figure \ref{fig:logMd_logMs_z0}, we bin the TNG100 sources by stellar mass and calculate the mean and standard error on the mean for galaxies in a given bin, as long as there are at least 50 sources in the bin.  

We see that, broadly, the dust mass estimates from local studies agree with the $z=0$ galaxies from TNG100 (Figure \ref{fig:logMd_logMs_z0}). However, it is worth noting that dust masses for the $K$-band-selected HRS, and the average dust mass in each stellar mass bin for the TNG100 simulations (which agree well with each other) are a little higher compared to the larger, but optically-selected, H-ATLAS sample from \citet{Beeston2018}. The offset between our mean TNG100 values and the mean values for \citet{Beeston2018} is $\sim$0.2 dex. The 850\,$\mu$m stacked sample from S2COSMOS lies somewhere between the HRS and \citet{Beeston2018}. It may be possible that the small absolute offsets in average dust masses seen here are enhanced by the different selection effects and sampling issues of the observational studies. In particular, the optically-selected sample of \citet{Beeston2018} have average dust masses sitting slightly below the FIR-selected samples of H-ATLAS sources (see their Figure 10). Conversely, the TNG100 galaxy sample is not subject to such observational selection effects. The deviation could also be attributed to the methods used to calculate stellar and dust masses in the observational studies and the simulation data, which are not identical. However, qualitatively, the results for the datasets considered in Figure \ref{fig:logMd_logMs_z0} are reasonably consistent, despite the different techniques and samples used to estimate physical parameters. We again remind the reader that the agreement shown in Figure \ref{fig:logMd_logMs_z0} is, in part, by construction due to the choice of $\kappa_{500}$ (see Section \ref{sec:4.1_localDMF}). 

At log$_{10}(M_*/M_{\odot}) \sim$ 10.5 in Figure \ref{fig:logMd_logMs_z0}, there is a distinct population of TNG100 galaxies with a lack of dust (grey crosses). Since these are massive galaxies, low resolution is not likely to be the cause of such low dust masses. As can be seen in the top-right panel of Figure \ref{fig:logMd_logMs_z0}, these galaxies are devoid of neutral gas. These TNG100 galaxies are `quenched' galaxies that have had their gas supply disrupted due to Active Galactic Nuclei (AGN) feedback mechanisms. In IllustrisTNG, this stellar mass regime is the threshold at which kinetic mode feedback from AGN (in the form of black hole driven winds) becomes a dominant feedback effect (\citealt{Weinberger2017, Weinberger2018}; \citealt{Terrazas2020}; \citealt{Zinger2020}). 

In Figure \ref{fig:logMd_logMs_z0}, at lower stellar masses, there is a spur of galaxies with high dust masses, which is not seen in the observed samples. These galaxies also have high metallicities (Figure \ref{fig:logMd_logMs_z0}, bottom-right panel). 

Despite some small differences, the galaxies from TNG100 broadly follow the same $(M_*-M_d)$ distribution that we see in observations at $z=0$, showing that with the exception of the low-mass end of the IllustrisTNG DMF, our post-processing method for estimating the dust mass does a good job of predicting the dust properties of the local galaxy population. Having performed this low-redshift check, we now investigate whether the discrepancy between the observed and predicted evolution over the redshift range $0<z<0.5$ continues to higher redshifts.

\subsection{High Redshift Dust Mass Evolution} \label{sec:4.3_stacking_compare}

\begin{figure*}
	\includegraphics[width=\textwidth]{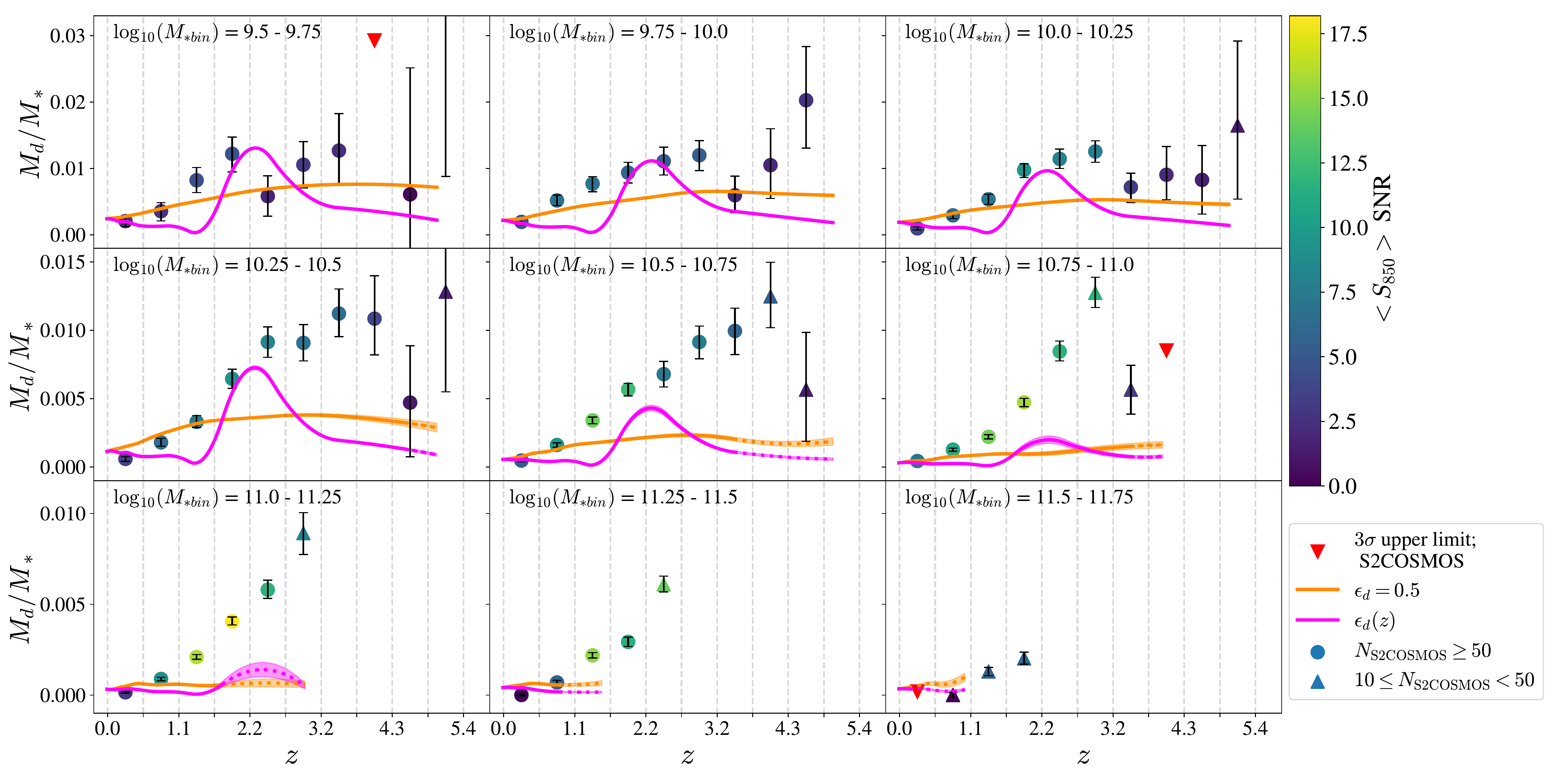}
    \caption{Dust-to-stellar mass ratios obtained from the stacked S2COSMOS fluxes (\textit{symbols}, \protect\citealt{Millard2020}) and TNG100 galaxies ({\it orange and magenta lines}). 
    \textit{Red triangles:} observed dust-to-stellar-mass ratios estimated using flux 3$\sigma$ upper limits. \textit{Circles:} observed dust-to-stellar mass ratios, where colour of point represents SNR of flux used to estimate dust masses, and there are at least 50 {\sc magphys}-COSMOS sources in ($M_* - z$) bin. \textit{Triangles:} number of {\sc magphys}-COSMOS sources in ($M_* - z$) bin, $N_{\rm S2COSMOS}$, between 10 and 50. 
    {\it Orange line}: dust-to-stellar-mass ratio for TNG100 galaxies assuming a constant $\epsilon_d$=0.5 (quadratic interpolation, see Figure \ref{fig:TNG_evolution_interp}). The {\it magenta line} shows instead the results when a redshift-dependent relation for $\epsilon_d(z)$ is assumed (\protect\citealt{Inoue2003} for $z<2$ and \protect\citealt{Vogelsberger2020} for $z \geq 2$). Note the sharp transition at $z \sim 1.7$.
    \textit{Solid line:} at least 50 TNG100 sources in($M_* - z$) bin. \textit{Dashed line:} number of TNG100 sources in ($M_* - z$) bin, $N_{\rm TNG}$, between 10 and 50. }
    \label{fig:stacked_TNG_evolution}
\end{figure*} 

After calculating the dust masses for each galaxy in each TNG100 snapshot (see Section \ref{sec:3.2_dustmasses_TNG_eqn}), we bin the galaxies into the same stellar mass bins used in our stacking analysis. For a given ($M_* - z$) bin that has at least ten sources, we calculate the mean of the dust-to-stellar-mass ratio $(M_d/M_*)$. The errors on the calculated dust-to-stellar-mass ratios are taken to be the standard error on the mean. Since the TNG100 snapshots are at different redshifts to the redshift bins used in our stacking analysis, we perform quadratic interpolations to the resulting TNG100 dust-to-stellar-mass ratios in each stellar mass bin, to allow a comparison with our stacking data. Here, we focus on comparisons between the observations and simulations when considering a constant $\epsilon_d = 0.5$. In Section \ref{sec:5.1_caveats_assumps}, we return to this comparison and consider the impact of an evolving $\epsilon_d(z)$ with redshift.

Figure \ref{fig:stacked_TNG_evolution} shows a comparison of the model predictions and observations. The model predicts dramatically less evolution than the observations (Figure \ref{fig:TNG_evolution_interp} shows just the TNG100 predictions with a scale chosen to highlight details of the predicted evolution more clearly).
At low stellar masses both the simulation and observations agree that the peak value of ($M_d/M_*$) is at $z \sim 3$, but TNG100 predicts much less evolution than is seen (Figure \ref{fig:stacked_TNG_evolution}). It is interesting to note that the peak in the dust-to-stellar-mass ratio of galaxies is a little before the peak of the star formation history in the Universe. At low redshifts, the dust-to-stellar-mass ratios estimated for the TNG100 galaxies agrees well with the dust-to-stellar-mass ratios calculated in the stacking analysis for the observed galaxies, in all stellar mass bins, but we remind the reader that the agreement at low redshifts is by construction (see Section \ref{sec:4.1_localDMF}).

The simulation and the observations agree that at low redshift the galaxies with high stellar masses generally contain less dust than those with lower stellar masses. However, 
the difference between the predicted and observed evolution for the
high-mass galaxies ($> {\rm log}(M_{*,{\rm bin}}) \sim 10.5$) is dramatic (Figure \ref{fig:stacked_TNG_evolution}). The observed evolution is very large and continues out to the highest redshifts probed by our sample. In comparison, the simulated galaxies still have very small amounts of dust at high redshifts. Figure \ref{fig:stacked_TNG_evolution} shows that the difference in dust-to-stellar-mass ratio evolution between the simulated and observed data is highest at the highest stellar masses - except for the highest stellar mass bin, where a lack of galaxies in both datasets prohibits meaningful conclusions on the evolution of dust-to-stellar-mass ratio to be drawn.

In the next Section, we explore reasons to explain the conflicting dust mass evolution shown between observations and simulations.

\section{Discussion} \label{sec:5_discussion}
\subsection{Caveats and assumptions}\label{sec:5.1_caveats_assumps}
Firstly, we consider the effect of the fundamental assumptions we have made in this work. Two big assumptions we have made are that the following parameters are time invariant; i) $\kappa_{500}$, and ii) $T_d$. 

There is no obvious reason why we would expect $\kappa_{500}$ to evolve with redshift, and we would need $\kappa_{500}$ to have increased by a factor of $\sim$5 by $z=3.5$ for galaxies with stellar masses ${\rm log}(M_{*,{\rm bin}}) = 10.5 - 10.75$ to bring the observations and the simulations into agreement. We are not aware of any predictions or observations showing such a large evolutionary effect in the properties of the dust itself, but we cannot rule this out.

In Equation \ref{Eq:Md_stack}, a higher dust temperature would lead to a lower calculated dust mass. In this work, it is important to remember that the dust temperature assumed in Equation \ref{Eq:Md_stack} is the \textit{mass-weighted} dust temperature, and therefore represents the temperature of the bulk of the ISM \citep{Millard2020}. As described in \cite{Millard2020}, mass-weighted dust temperature depends on the mean radiation energy density to the power of $\sim$1/6 \citep{Scoville2016}, and so vast differences in the ISM environment over the history of the universe would be required to significantly change the value of $T_d$. Thus far, there is little evidence to support such differences. Further, simulations of $z = 2-6$ galaxies by \cite{Liang2019} showed that the mass-weighted dust temperature of galaxies evolves little over this redshift range, and that a mass-weighted dust temperature of 25\,K is a not an unreasonable assumption. One factor to consider is the increasing temperature of the cosmic microwave background (CMB) with redshift \citep{daCunha2013} which can become non-negligible at $z>4$.  Using Equation~10 from \cite{daCunha2013}, the expected rise in temperature due to the CMB for dust grains with $T_d=25\,$K at $z=0$ would amount to only 0.4\,K at $z=5$. 

\begin{figure}
	\includegraphics[width=\columnwidth]{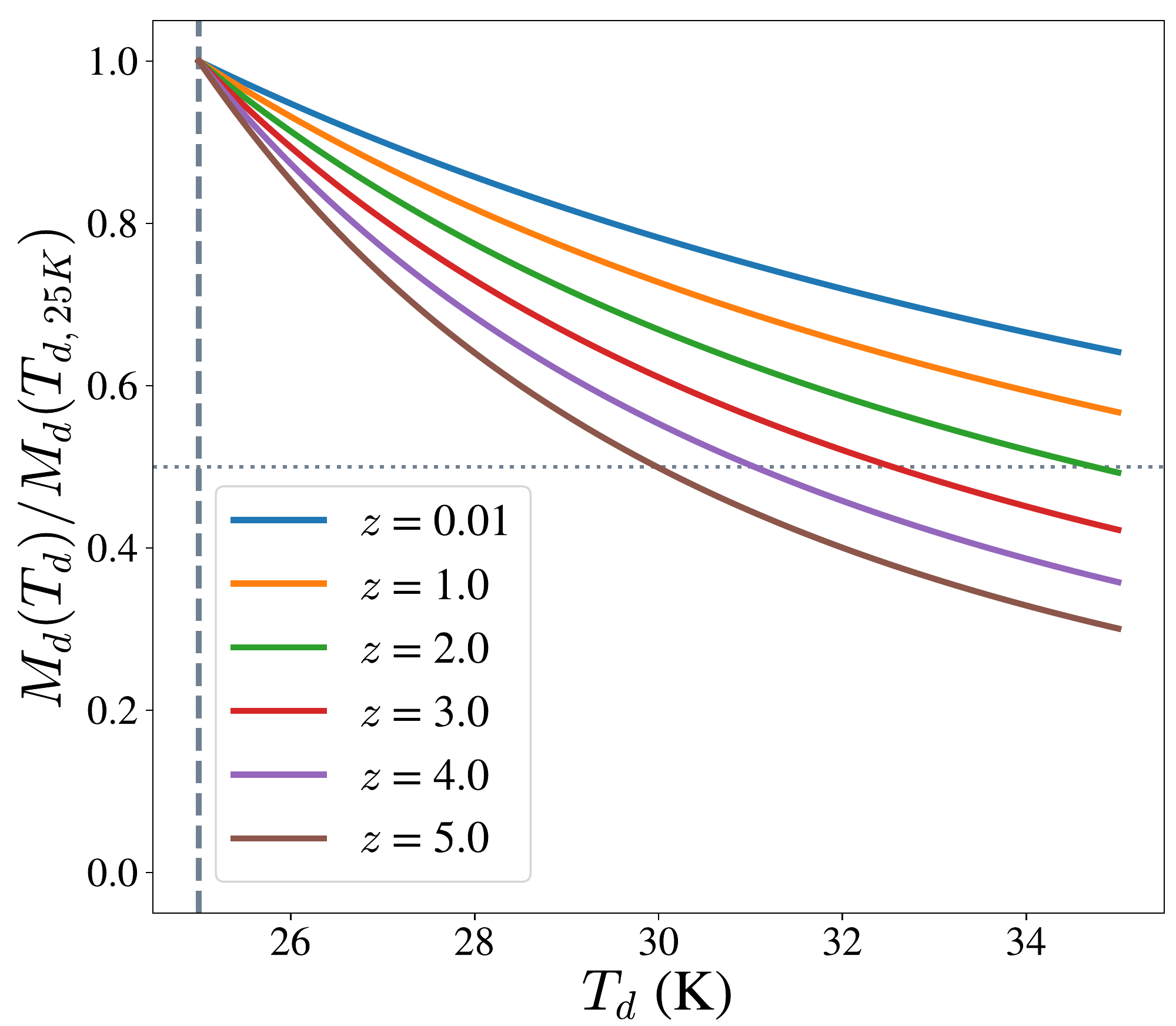}
    \caption{The variation of dust mass with assumed mass-weighted dust temperature for discrete redshifts. \textit{Vertical dashed line:} the mass-weighted dust temperature assumed in this study (25\,K). \textit{Horizontal dotted line:} 50 per cent reduction in dust mass.}
    \label{fig:changingTd}
\end{figure}

Figure \ref{fig:changingTd} shows how the observed dust masses change for discrete redshifts depending on the assumed dust temperature. We can see that even for an extreme increase in the mass-weighted dust temperature to 35\,K, the corresponding decrease in the estimated dust masses is not enough to to reconcile the differences in the dust-to-stellar-mass ratio at high redshifts as shown in Figure \ref{fig:stacked_TNG_evolution}.

In Equation \ref{Eq:scale_kappa}, we assumed a value of the dust emissivity spectral index, $\beta$, of 1.8, robustly determined by Galactic ISM studies using sub-mm/mm emission from Planck \citep{Planck2015}. This is also in agreement with a long-wavelength study of 29 Submillimetre Galaxies (SMGs; rapidly star-forming, highly dust-obscured massive galaxies typically found at $z \sim 2-4$) at $z \sim 2.7$ by \cite{Chapin2009}, who find an average $\beta$=1.75. Since the observational dust masses are inversely proportional to $\kappa_{\lambda_e}$, the choice of $\beta$ directly impacts the resulting dust masses (Equation \ref{Eq:Md_stack}).

Observational studies on extragalactic scales find values of $\beta$ = 1-2 (e.g. \citealt{Hildebrand1983}; \citealt{Dunne2001}; \citealt{Chapin2009}; \citealt{Clements2010}). Although these studies typically converge towards the idea that a value of $\beta$ towards the upper end of this range is appropriate for dust in galaxies, it is worth examining the impact the choice of a lower $\beta$ would have on the calculated observational dust masses. Therefore, as an  example, we consider the impact on our resulting dust masses had we chosen a significantly lower value of $\beta = 1.3$. 

\begin{figure}
	\includegraphics[width=\columnwidth]{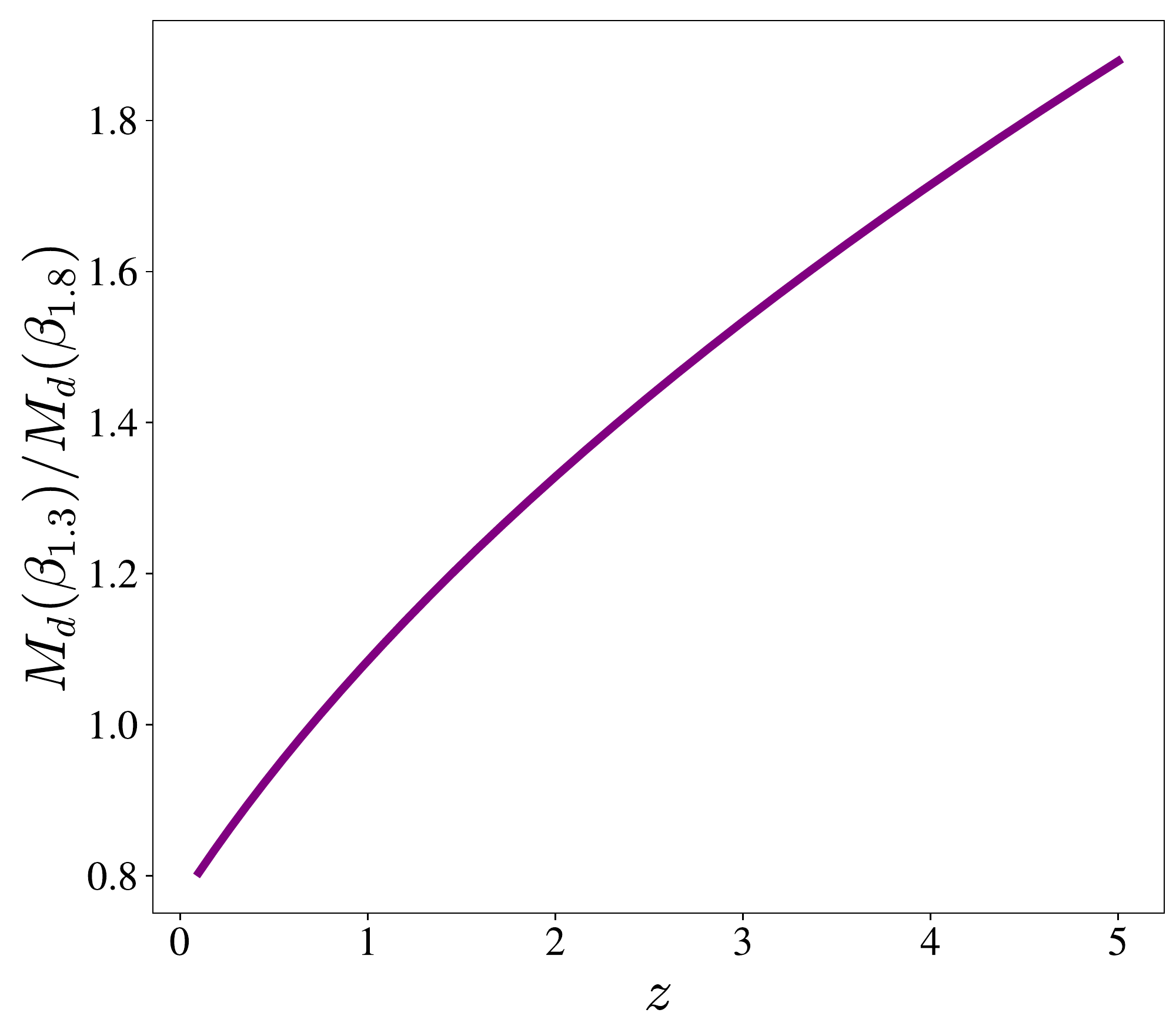}
    \caption{The variation of dust mass with redshift for two assumed values of the dust emissivity spectral index, $\beta$.}
    \label{fig:changing_beta}
\end{figure}

Figure \ref{fig:changing_beta} shows the variation in observational dust mass with redshift for our assumed value of $\beta=1.8$ and our test value of $\beta = 1.3$. For $z \lesssim 0.7$, we see that a lower value of $\beta$ would lead to lower dust masses compared to dust masses estimated using $\beta=1.8$. However, beyond this redshift, we see that a lower value of $\beta$ would lead to higher dust mass estimates, only exacerbating the difference between observations and the model seen in Figure \ref{fig:stacked_TNG_evolution}.

An important assumption that we have made in estimating the dust masses for TNG100 galaxies is that the fraction of metals locked up in dust grains, $\epsilon_d$, has a constant
value, (Section~\ref{sec:3.2.1_constant_epsilon_d}).

By setting its value to 0.5 we have left very little room to explain the discrepancy in Figure \ref{fig:stacked_TNG_evolution} by evolution in $\epsilon_d$. Even incorporating all the metals in a high-redshift galaxy in dust grains would only increase the value of $\epsilon_d$ to one, which would only double the high-redshift dust masses, much less than the discrepancy in Figure \ref{fig:stacked_TNG_evolution}. 
Such a scenario is not physical since the difference between the cosmic abundances of elements and the likely chemical composition of dust grains mean that large proportions of plentiful elements such as oxygen must remain in the gas phase (\citealt{Meyer1998}).
However, as a final check, we use the relationship of $\epsilon_d$ with $z$ from \citet{Vogelsberger2020} (valid for $2\leq z<9$) and values from \citet{Inoue2003} for $0<z<2$ (their Figure~6) and re-run the analysis in Figure~\ref{fig:stacked_TNG_evolution}. As before, we perform a quadratic interpolation to the discreet TNG100 results. We note that for the variation of $\epsilon_d$ with redshift, this results in a sharp transition at around $z \sim 1.7$ as we move from the $\epsilon_d(z)$ values of \cite{Inoue2003} used for the $z=1.5$ TNG100 file, to the $\epsilon_d(z)$ values of \cite{Vogelsberger2020} used for the $z \geq 2$ TNG100 files.

As shown in Figure \ref{fig:stacked_TNG_evolution}, this variation in $\epsilon_d$ with redshift provides a moderately better agreement with the observations at intermediate redshifts ($z \sim 2$) for the three lowest stellar mass bins, where the discrepancy between the observations and simulations (assuming a constant $\epsilon_d$) is the least. For stellar mass bins ${\rm log}(M_{*,bin}) = 10.25-10.75$, there is a good agreement between the observations and simulations at intermediate redshifts when using $\epsilon_d(z)$. However, this agreement requires a value of $\epsilon_d(z=2)$ close to 1, which, as previously explained, is physically unlikely, and is still not enough to explain the discrepancy at the highest redshifts ($z > 2.5$). Out to higher redshifts, and for the higher stellar mass bins, this evolving $\epsilon_d(z)$ does not fix the dichotomy in the evolution.

It is also worth noting that the evolving $\epsilon_d(z)$ leads to worse agreement between the observations and simulations at low redshifts ($z<2$). The relation from \cite{Inoue2003} gives a $\epsilon_d(z)$ that decreases with redshift. Therefore, the lack of agreement with a $\epsilon_d(z)$ that decreases with redshift, combined with the agreement with the high $\epsilon_d(z)$ from \cite{Vogelsberger2020} at intermediate redshifts and at low stellar masses, possibly indicates a requirement for an increasing $\epsilon_d(z)$ from $z=0$ out to $z \sim 2-2.5$, where the star formation rate density of the Universe peaks \citep{Daddi2007,Madau2014}. Figure~\ref{fig:stacked_TNG_evolution} also indicates that any variation in $\epsilon_d$ would need to have some dependence on galactic stellar mass. However, this is assuming that the COSMOS stacked observations are correct, and assuming that the only parameter that needs to be changed is $\epsilon_d$. 

Another parameter for which $\epsilon_d$ could vary is metallcity. To remove the discrepancy in Figure~\ref{fig:stacked_TNG_evolution} would require a value greater than the currently assumed value of 0.5 at higher redshifts. Figure~\ref{fig:dust_metals} shows that any variation in $\epsilon_d$ with metallicity (eg \citealt{Remy-Ruyer2014}) tends to reduce the value below 0.5, locking even less 
metals into dust. Therefore an $\epsilon_d$ that varies with metallicity cannot explain the discrepancy.

One place where a change in $\epsilon_d$ may be important is in explaining the discrepancy between the observed and predicted low-redshift DMFs in Figure \ref{fig:lowz_DMF} for the \emph{lowest dust mass} galaxies. At low dust masses the TNG100 DMF is higher than the observed DMF. The galaxies with these
low dust masses have lower stellar masses than the galaxies in our COSMOS sample
and are therefore more likely to resemble the low-metallicity systems for which
there is some evidence that $\epsilon_d$ is lower than in the Milky Way (Section~\ref{sec:3.2.1_constant_epsilon_d}). 
Lowering $\epsilon_d$ by a factor of $\sim 2$ would
be enough to resolve the discrepancy in the DMF at low dust mass, in agreement with the scatter in $\epsilon_d$ we see in galaxies in the local Universe (Figure~\ref{fig:dust_metals}). However, this cannot be responsible for the discrepancy in \emph{the evolution} observed in the low redshift DMFs (Figure~\ref{fig:lowz_DMF}), since models that predict a variation in $\epsilon_d$ over $0<z<0.5$ produces a difference that is too small to account for the evolution observed in \citet{Dunne2011}. For example, \citet{Inoue2003} find a variation of only a factor of 2 in $\epsilon_d$ over this range (Figure~\ref{fig:dust_metals}).

\subsection{IllustrisTNG: Satellites vs. centrals}\label{sec:5.2_sats_cents}
Previous studies examining the neutral gas content of IllustrisTNG galaxies at $z=0$ have found that environment is an important influencing factor. For example, \cite{Stevens2019} showed that $z=0$ TNG100 satellite galaxies are typically a factor of $>$3 poorer in HI than a central galaxy of the same stellar mass. In addition, they showed that the inherent neutral gas fractions of TNG100 satellite galaxies show a dip at around ${\rm log}(M_*/M_{\odot}) \sim 10.75$. They attribute the lack of gas in satellite galaxies to i) gas lost via AGN feedback (satellites are less able to reattain any gas lost due to their lack of gravitational influence); and ii) stripping and quenching. Further, \cite{Diemer2019} found a population of TNG100 galaxies, the vast majority of which are classified as satellites, that were devoid of gas. These galaxies were found to largely live in crowded environments, and so have likely been stripped of their gas by a larger halo host. 

\begin{figure*}
    \includegraphics[width=\textwidth]{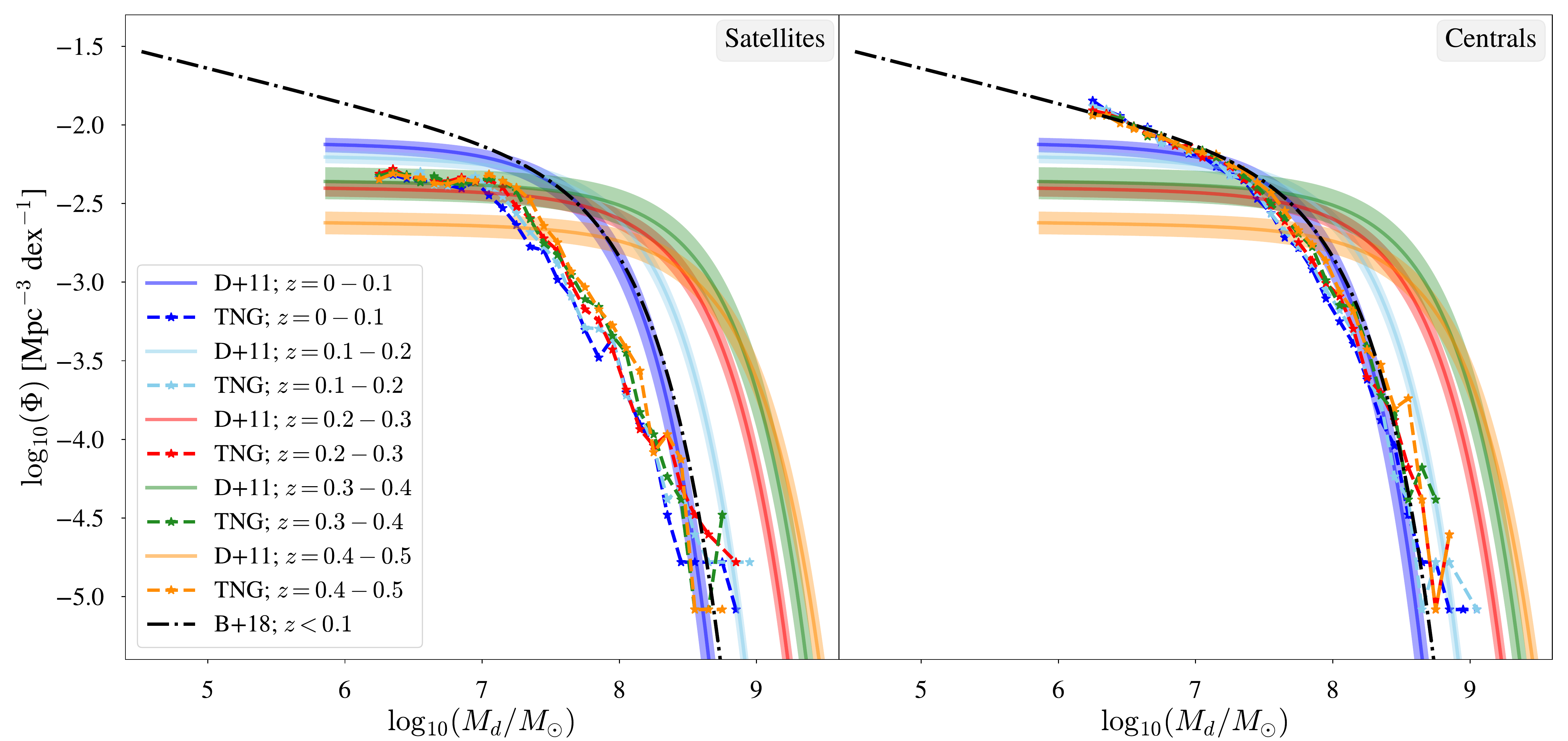}
    \caption{Low redshift DMFs for IllustrisTNG galaxies {\it (stars and dashed lines)} and local DMFs from \protect\cite{Beeston2018} {\it (black dash-dot line)} and \protect\cite{Dunne2011} {\it (solid colour lines)}. The IllustrisTNG sources have been split into two populations: satellite {\it (left)} and central galaxies {\it (right)}.Literature DMFs are the same as in Figure \ref{fig:lowz_DMF}, and as before, have been scaled to IllustrisTNG cosmology and the values of the dust-mass opacity coefficient, $\kappa_{500}$, used in this paper.}
    \label{fig:TNG_DMF_sats_cents}
\end{figure*}

\begin{figure*}
    \includegraphics[width=\textwidth]{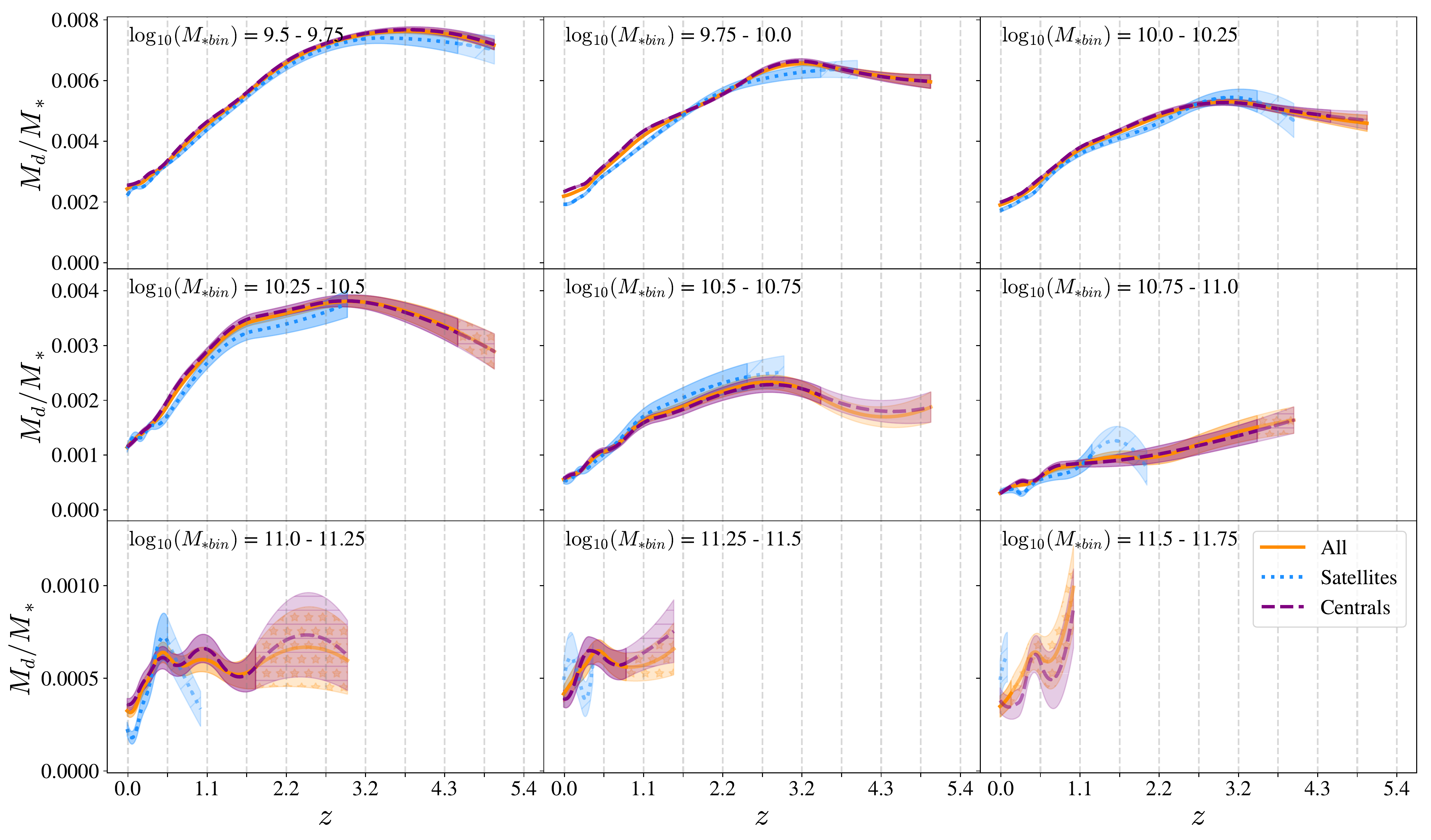}
    \caption{Dust-to-stellar-mass ratios for TNG100 galaxies plotted against 
    redshift. \textit{Orange solid line:} all galaxies. \textit{Purple dashed line:} central galaxies. \textit{Blue dotted line:} satellite galaxies. The paler, hatched areas represent data where the number of TNG100 sources in a given $(M_*-z)$ bin is between 10 and 50. Solid, non-hatched areas represent data where the number of sources in a given $(M_* - z)$ bin is at least 50. The lines are quadratic interpolations to discrete data points calculated at the redshifts of the TNG100 datafiles (Table \ref{tab:TNGnumbers}). The orange line shown here is the same as in Figures \ref{fig:TNG_evolution_interp} and \ref{fig:stacked_TNG_evolution}.}
    \label{fig:TNG_MdMs_all_sats_cents}
\end{figure*}

It is therefore not unreasonable to consider that such unusual gas-poor satellite galaxies may be masking more rapid dust mass evolution in TNG100 galaxies than is shown by the collective population. To test this, we split our simulated galaxy sample into satellites and centrals\footnote{According to the {\tt Subfind} classification} and separately examine the dust mass evolution for these two galactic populations (Figures \ref{fig:TNG_DMF_sats_cents} and \ref{fig:TNG_MdMs_all_sats_cents}).

Figure \ref{fig:TNG_DMF_sats_cents} shows the DMFs for the satellite and central TNG100 galaxies. Considering first the central galaxies (Figure \ref{fig:TNG_DMF_sats_cents}; \textit{right}), we can see that the $z=0$ TNG100 DMF agrees remarkably well with the $z=0$ DMF from \cite{Beeston2018} at low and high dust masses, although the knee of the TNG100 DMF sits slightly lower than the observed dust mass function. This shows that at $z=0$, the dust masses of TNG100 central galaxies are well-modelled by the simulations plus our post-processing recipe. 

Now examining the DMFs of the satellite galaxies (Figure \ref{fig:TNG_DMF_sats_cents}; \textit{left}), we see that in comparison to observations, TNG100 satellite galaxies are particularly devoid of dust. This is not unexpected, considering the lack of neutral gas in satellite galaxies found by previous studies (e.g. \citealt{Diemer2019}; \citealt{Stevens2019}). 

A comparison of Figures \ref{fig:lowz_DMF} and \ref{fig:TNG_DMF_sats_cents} shows that the excess number of low dust mass sources in the total DMF can be attributed to satellite galaxies with low gas content, probably due to a combination AGN feedback driving gas out of the gravitational influence of satellite galaxies, ram-pressure stripping, and quenching (\citealt{Stevens2019}). 

However, the most striking and important feature of Figure \ref{fig:TNG_DMF_sats_cents} is that splitting the TNG100 galaxies sample into satellite and centrals \textit{does not solve the lack of dust mass evolution} over the redshift range $0<z<0.5$ in TNG100 galaxies as compared to observations. Both satellites and central galaxies show a lack of evolution. This indicates that the lack of dust mass evolution in TNG100 galaxies can be attributed to global simulation properties, rather than secondary effects that manifest due to environmental effects, for example.

Next, we explore the dust mass evolution of satellite and central IllustrisTNG galaxies out to high redshifts. Figure \ref{fig:TNG_MdMs_all_sats_cents} shows the evolution in the dust-to-stellar-mass ratio of TNG100 galaxies for galaxies with stellar masses ${\rm log}(M_*/M_{\odot}) \geq 9.5$. As shown in Figure \ref{fig:TNG_MdMs_all_sats_cents}, the dust-to-stellar-mass ratio evolution of satellite and central galaxies are largely similar, particularly for stellar masses ${\rm log}(M_*/M_{\odot}) \leq 11.0$. Above this stellar mass limit (i.e. for galaxies in the three highest stellar mass bins), the dust-to-stellar-mass ratio evolution of satellite galaxies is a little different to that of centrals and the combined sample. However, we caution against inferring too much here, since there are few satellite galaxies in these bins, and the results suffer from low number statistics. 

It is interesting to note that in some stellar mass bins, satellite galaxies have slightly higher dust-to-stellar-mass ratios compared to the central galaxies (e.g. ${\rm log}(M_{\rm *,bin}) = 10.5-10.75$). In a given stellar mass bin, the average stellar mass of satellite galaxies is slightly lower than the average stellar mass of central galaxies. For galaxies with a similar dust mass, this therefore acts to marginally increase the dust-to-stellar-mass ratio for satellite galaxies. 

Further, the similarity in the dust-to-stellar-mass ratio evolution of the two populations with redshift might seem strange when we consider the difference between the DMFs for satellite and central galaxies (Figure \ref{fig:TNG_DMF_sats_cents}). However, in Figure \ref{fig:TNG_MdMs_all_sats_cents}, we are probing the highest stellar mass systems, and there is little difference between the high mass ends of the DMFs for the satellite and central galaxies.

Similar to Figure \ref{fig:TNG_DMF_sats_cents}, the most striking and important feature of Figure \ref{fig:TNG_MdMs_all_sats_cents} is that splitting the sample into satellite and central galaxies \textit{does not solve the lack of dust mass evolution} over the redshift range $0<z<5$ in TNG100 galaxies as compared to observations. 

We now explore global properties of the IllustrisTNG simulation in an attempt to explain the lack of evolution in the dust mass of galaxies over cosmic time as compared to observations. Specifically, in the next sections, we examine metal enrichment prescriptions (Section \ref{sec:5.3_metal_prescriptions_TNG}), the neutral gas content of galaxies (Section \ref{sec:5.4_gas_mass_lit}), and feedback mechanisms (Section \ref{sec:5.5_feedback_mech}).

\subsection{IllustrisTNG metal enrichment prescriptions}\label{sec:5.3_metal_prescriptions_TNG}
As described in \cite{Torrey2019}, in the IllustrisTNG simulations, metals are returned to the ISM by both supernovae explosions and asymptotic giant-branch (AGB) stars. In the TNG100 simulations (the data used in this study), individual stars are not resolved; star particles have masses of $m_* \sim 10^6 M_{\odot}$, and these star particles are assumed to encapsulate a \cite{Chabrier2003} initial mass function {IMF} \citep{Torrey2019}. At each timestep in the simulation, stellar lifetime tables \citep{Portinari1998} determine which stars leave the main sequence. Upon the death of stars, mass return and metal yield tables are used to determine the mass of metals returned to the ISM (\citealt{Nomoto1997}; \citealt{Portinari1998}; \citealt{Kobayashi2006}; \citealt{Karakas2010}; \citealt{Doherty2014}; \citealt{Fishlock2014}). The returned metal mass is spread over the nearest 64 gas cells. 

In IllustrisTNG, galactic winds are probabilistically estimated, based on local SFRs (\citealt{Torrey2019}; see \citealt{Pillepich2018b} for details). They are assigned a lower metallicity than the ISM of a galaxy ($Z_{\rm wind} = 0.4Z_{\rm ISM}$), to encapsulate the dilution of metal-rich gas by metal-poor gas as the winds drive the metal-rich gas away from their origin point.

In their investigations into the IllustrisTNG mass-metallicity relation, \cite{Torrey2019} found that a significant proportion of the metals of simulated galaxies lie outside of the cool ISM, where we would expect dust to form. They also found that higher-mass IllustrisTNG galaxies are less efficient at retaining metals compared to lower-mass counterparts. We see this reflected in Figure \ref{fig:stacked_TNG_evolution}, where the discrepancy between the evolution of dust shown in the observed and simulated galaxies is greater in the bins with higher stellar masses. Therefore, one possible explanation of the low dust masses in the model is that TNG100 ejects too many metals from the cool ISM.

\subsection{IllustrisTNG gas mass evolution}\label{sec:5.4_gas_mass_lit}
Since we derive dust masses for IllustrisTNG galaxies using the neutral gas mass, the lack of evolution in the dust masses might stem from a lack of evolution in the mass of the cool ISM.

Several studies have compared the cold gas mass content of IllustrisTNG galaxies with observations of the local universe. \cite{Stevens2019} found that the neutral gas fractions of TNG100 galaxies at $z=0$ agree well with observations, although the most massive central galaxies ($M_* > 10^{10.7} M_{\odot}$) are somewhat too gas rich, and the gas content of satellite galaxies sharply dips at masses $M_* \sim 10^{10.5} M_{\odot}$. Similarly, \cite{Diemer2019} showed that the HI and $\rm H_2$ mass fractions ($M_{\rm HI+H_{2}}/M_*$) of TNG100 galaxies at $z=0$ largely agree with observations, with some discrepancies in the H$_2$ mass fraction for the most massive galaxies ($M_* > 2 \times 10^{10} M_{\odot}$), which they found to be at least a factor of 4 lower compared to observations. However, \cite{Diemer2019} also found that the $z=0$ TNG100 HI and H$_2$ mass functions largely agree with observations. This result is echoed in \cite{Dave2020}, although they do note that the HI and H$_2$ mass functions of TNG100 are slightly too high at the high gas mass end. 

It is worth noting that in the aforementioned studies (\citealt{Dave2020}; \citealt{Diemer2019}; \citealt{Stevens2019}), comparisons to observational data are not straight-forward - measuring the neutral gas content of galaxies accurately is notoriously difficult. For example, the CO-to-H$_2$ conversion factor, used to estimate the mass of molecular gas from observations of the CO line, is uncertain and varies from galaxy type to galaxy type (e.g. \citealt{Bolatto2013}; \citealt{Bethermin2015}; \citealt{Genzel2015}; \citealt{Scoville2016}; \citealt{Popping2019}), and galaxies can contain CO-dark molecular gas (\citealt{Bolatto2013}). When considering these observational limitations, these studies have shown that in the local universe, the neutral gas content of IllustrisTNG galaxies agrees well with observations.

Accurately measuring the gas content of galaxies at high redshifts is even more difficult. For example, although 21cm radio emission is efficient at detecting HI gas in galaxies at $z=0$, it is currently hard to detect 21cm radiation at $z > \sim 0.2$ (\citealt{Catinella2010}). Typically beyond the local universe, gas mass estimates are limited to molecular gas estimates, usually derived using CO as a tracer (e.g. \citealt{Solomon2005}; \citealt{Coppin2009}; \citealt{Tacconi2010}; \citealt{Casey2011}; \citealt{Bothwell2013}; \citealt{Carilli2013}; \citealt{Tacconi2013}; \citealt{Combes2018}).

Despite these difficulties, there have been several attempts to compare observations of the evolution in the cool gas in galaxies with predictions by models. Interestingly, \citealt{Dave2020} found that the IllustrisTNG HI mass function shows a negative evolution with redshift, and that the H$_2$ mass function is mostly unevolving with redshift (out to $z < 2$), implying that the gas content of IllustrisTNG galaxies varies little with redshift. They found that the evolution in HI and H$_2$ was much weaker than predicted by two other simulations: EAGLE \citep{Dave2019} and {\sc simba} \citep{Schaye2015}. Similar results were found by \cite{Diemer2019}, who found a very weak evolution in the abundance of HI in TNG100 galaxies between $z=1-4$, and a moderate evolution in the abundance of H$_2$ over the redshift range $z=0-4$, with a peak increase of a factor of 2-3 at around $z=2$. Recently, \cite{Popping2019} showed that the H$_2$ masses predicted by TNG100 for galaxies at $z>1$ were a factor of 2-3 lower than observations (when comparing to observations using a standard CO-to-H$_2$ conversion factor).

All these studies suggest that TNG100 galaxies beyond $z \sim 1$ are lacking in neutral gas, and that there is little evolution in the gas content of the simulated galaxies with redshift. Here, by using post-processing methods to compare the results of IllustrisTNG with the evolution of dust in galaxies over cosmic time, we get a similar result. The lack of evolution in the dust mass in TNG100 compared to observations could therefore be a reflection of the lack of evolution of the neutral gas content in the simulations that has been seen previously.  In the next Section, we speculate as to what may be driving the different gas evolution, and therefore dust evolution, of the IllustrisTNG galaxies.

\subsection{IllustrisTNG feedback mechanisms}\label{sec:5.5_feedback_mech}
Whether the lack of evolution in the dust predicted by TNG100 is entirely linked to the lack of evolution in the gas content (Section \ref{sec:5.4_gas_mass_lit}) or whether it is also linked to the ejection of metals, it is clearly linked to the model ejecting too much material outside a galaxy. It therefore seems likely to be caused by excessive galaxy outflows or feedback.

As discussed in Section \ref{sec:4.2_local_dust_mass}, kinetic feedback from AGN in the form of black hole driven winds becomes the dominant feedback process over other feedback methods (e.g. stellar feedback) above galactic stellar masses of log$_{10}(M_*/M_{\odot}) \sim 10.5$, particularly at late times. Kinetic AGN feedback in IllustrisTNG manifests in a two-fold manner - firstly, it acts to expel star-forming gas from galaxies. Secondly, at these stellar masses, the AGN feedback increases the gas cooling time, preventing radiative cooling and future gas accretion (\citealt{Terrazas2020}; \citealt{Zinger2020}). As noted in \citealt{Dave2020}, at earlier epochs, it is the second of these processes, thermal feedback, which is most
important, reducing the amount of cold gas in high-redshift galaxies.
The fact that the disagreement between the model predictions and the observations is greatest at the high stellar masses is at least circumstantial evidence that the explanation of the discrepancy might be too vigorous AGN feedback. 

\subsubsection{Insights from Illustris}
Hayward et al. (2020) have recently investigated whether Illustris and IllustrisTNG can replicate the numbers of the rare luminous Submillimetre
Galaxies (SMGs), using the star-formation rates predicted by the simulations
and post-processing estimates of the dust mass. They found that the predicted number counts of SMGs in Illustris was largely comparable with observations, but that IllustrisTNG notably lacked SMGs. Further investigation led to the revelation that the identified IllustrisTNG SMGs are particularly dust-poor, with some SMGs (${\rm log}(M_*/M_{\odot}) \approx 11$) at $z \sim 2$ in Illustris having a factor of three higher dust content than the IllustrisTNG counterparts. \citealt{Hayward2020} ultimately traced this dichotomy to a lack of gas, as opposed to a lack of metals, in these IllustrisTNG galaxies as compared to Illustris, driven by changes to the feedback model between Illustris and IllustrisTNG - either stellar feedback outflows, and/or AGN feedback. 
The changes to the feedback model were made to make quenching more efficient and so
bring IllustrisTNG galaxies more inline with the observed $z=0$ colour bimodality (\citealt{Weinberger2017, Weinberger2018}; \citealt{Nelson2018}). However, \cite{Hayward2020} point out that these changes seem to have quenched $z \sim 2-3$ galaxies too early, leading to a direct lack of massive dusty galaxies in IllustrisTNG as compared to observations.

Whilst the results of \cite{Hayward2020} are not directly comparable to this study due to their specific galactic population selection, the general idea that changes to feedback in IllustrisTNG have consequently quenched galaxies too soon or too frequently is inline with the results of this study. If the inability of IllustrisTNG plus our post-processing recipe to match the strong evolution in the dust masses is not caused by cosmic evolution in the properties of the dust grains themselves, it seems most likely that is caused by a lack of cool gas in the high-redshift TNG galaxies, possibly caused by over-enthusiastic AGN feedback.

\subsection{Comparison with EAGLE}
Recently, \cite{Baes2020} derived DMFs out to $z=1$ for galaxies from the EAGLE simulation. Dust masses were estimated using modified blackbody fits to synthetic infrared luminosities at FIR wavelengths (160$\mu$m, 250, 350 and 500$\mu$m) generated using a post-processing 3D radiative transfer procedure (\citealt{Baes2011}; \citealt{CampsBaes2015}; \citealt{Camps2016, Camps2018}). Similarly to the results of this study, in the local universe ($z<0.1$), \cite{Baes2020} found that they could reproduce the shape and normalization of the DMF fairly well, getting very good agreement with the DMFs found by \cite{Dunne2011} and \cite{Beeston2018} for dust masses $M_d < 2 \times 10^7 M_{\odot}$ but predicting too few galaxies at higher masses. 

When examining the evolution of the EAGLE DMF up to $z=1$, \cite{Baes2020} found only a mild evolution in the characteristic mass of the modified schechter functions fitted to the DMFs, and very little evidence for density evolution. They did, however, find a fairly good agreement with the weak evolution in the cosmic dust mass density found by \cite{Driver2018} for this redshift range.

It is interesting that both the EAGLE and IllustrisTNG simulations show a similar lack of evolution in the DMF as compared to observations, despite two different methods to estimate dust masses in post-processing, which shows that cosmological hydrodynamical simulations are still limited in their ability to reproduce the strong evolution seen in the global properties of dust in galaxies.

\section{Conclusions}
In this work, we have compared the observed evolution of the dust in galaxies with the predictions from a model. As our observations, we used previous estimates of the dust-mass function (DMF) over the redshift range $0<z<0.5$ and our own estimates of the mean ratio of dust mass to stellar mass over the redshift range
$0<z<5$. We made our predictions using the  cosmological hydrodynamical simulation IllustrisTNG (TNG100), using a simple post-processing recipe in which half
the metals in the cool ISM are locked up in dust grains.

\begin{itemize}
\item We created `local' DMFs based on galactic dust mass estimates for TNG100 galaxies out to $z=0.5$, and compared these to previously observed DMFs over the same redshift range. We find the DMFs from TNG100 show little evolution, which is in stark contrast with the strong evolution seen in the empirical DMFs. 

\item We find that the observed galactic population show a strong evolution in the dust-to-stellar-mass ratio ($M_d/M_*$) up to $z=5$. For lower stellar mass bins (${\rm log}(M_{*,{\rm bin}}) < 10.75$), we find that the dust-to-stellar-mass ratio peaks at $z=3-4$, but we cannot locate the peak for the high-mass galaxies because of the lack of high-mass galaxies at high redshifts. 
 We find that galaxies with a high stellar mass generally contain less dust than
 galaxies with a low stellar mass at the same redshift. 

\item The model predicts much weaker evolution than the observations
over the redshift range $0<z<5$. At lower stellar masses (${\rm log}(M_{*,{\rm bin}}) < 10.75$) the predicted evolution has a similar redshift dependence
to the observed evolution but that for all stellar masses
the strength of the evolution is much weaker than for the observations.

\item We split our simulated galaxies into two samples, satellites and centrals, and tested to see if the lack of dust mass evolution observed in the full sample was biased due to the dust content of one of these galaxy populations. We found that although satellite galaxies are typically more dust-poor than their central counterparts, both satellites and centrals show a similar lack of dust mass evolution as seen for the collective sample. Splitting the sample into satellite and central galaxies does not solve the lack of dust mass evolution over the redshift range $0 < z < 5$ in TNG100 galaxies as compared to observations.

\item We find that it is very difficult to bring the observed and model dust mass evolution into agreement by changes in the assumptions underpinning
our observations. The obvious ways of doing this are: i) a drastic evolution in the dust-mass opacity
coefficient with redshift; ii) a non-constant dust-to-metals ratio; iii) an extreme increase in the mass-weighted dust temperature with redshift; or iv) an extremely high value of $\beta \gg 2$.
The third and fourth of these are inconsistent with observations. Therefore the only possible ways, from the observational side, of making the
observations and theory consistent would be to assume that the properties
of the dust itself change drastically with redshift. This would require the dust-mass opacity coefficient to be much higher at high redshifts than at low redshifts. There is no
particular reason to think this should happen but we cannot rule this out.
We also show that the results of our study are robust against changes to our assumption about the fraction of metals bound up in dust grains: varying $\epsilon_d$ with metallicity cannot account for the discrepancy we observe. An $\epsilon_d$ that varies with redshift, however, can reduce, though not eliminate, the discrepancy if i) it increases with redshift ii) is a strong function of stellar mass and iii) is extremely high ($\geq 0.8$ would be needed in the higher redshift bins). 

\item Although determining the root cause of the discrepancies between simulations and observations is difficult, we attribute the differences to one or more of the following in the models: i) excessive galactic winds driving metals out of the cool ISM where we expect dust to form; ii) a lack of evolution in the neutral gas content of galaxies with redshift; iii) kinetic feedback from AGN expelling gas from galaxies.

\item We note that a lack of evolution in the dust content of galaxies as compared to observations is not limited to IllustrisTNG, but has also been in the $z<1$ DMFs calculated for the EAGLE simulation, despite using a different post-processing technique for estimating dust masses.

\end{itemize}

\noindent Previous studies of neutral gas (HI+H$_2$) have concluded that the neutral gas content in IllustrisTNG does not show sufficient evolution with redshift. However, at high redshifts, H$_2$ and (more so) HI are hard to measure. It is much easier to measure the dust masses of high-redshift galaxies than their gas masses. By using IllustrisTNG with a simple post-processing technique we have been
able to predict the dust masses of high-redshift galaxies. We find that the observed evolution is much stronger than the predicted evolution. If the discrepancy is not produced by cosmic evolution in the properties of interstellar dust itself, the most likely explanation seems to be that TNG does not predict strong enough evolution in the neutral gas content of galaxies.

\section*{Acknowledgements}
We thank Claudia Lagos, Dylan Nelson, and Annalisa Pillepich for helpful comments on the draft manuscript. We also thank Matthieu Bethermin for a productive and thorough referee report, which significantly improved this paper.
JSM, HLG and RAB acknowledge support from the European Research Council in the form of Consolidator Grant CosmicDust (ERC-2014-CoG-647939, PI H L Gomez). SAE and MWLS gratefully acknowledge the support of a Consolidated Grant
(ST/K00926/1) from the UK Science and Technology Funding Council (STFC). JSM would also like to thank N Parry and O Millard-Parry for their support. 

The James Clerk Maxwell Telescope is operated by the
East Asian Observatory on behalf of The National Astronomical Observatory of Japan; Academia Sinica Institute
of Astronomy and Astrophysics; the Korea Astronomy and
Space Science Institute; the Operation, Maintenance and
Upgrading Fund for Astronomical Telescopes and Facility
Instruments, budgeted from the Ministry of Finance (MOF)
of China and administrated by the Chinese Academy of Sciences (CAS), as well as the National Key R\&D Program
of China (No. 2017YFA0402700). Additional funding support is provided by the Science and Technology Facilities
Council of the United Kingdom and participating universities in the United Kingdom and Canada (ST/M007634/1,
ST/M003019/1, ST/N005856/1). The James Clerk Maxwell
Telescope has historically been operated by the Joint Astronomy Centre on behalf of the Science and Technology
Facilities Council of the United Kingdom, the National Research Council of Canada and the Netherlands Organization for Scientific Research and data from observations undertaken during this period of operation is used in this
manuscript. This research used the facilities of the Canadian
Astronomy Data Centre operated by the National Research
Council of Canada with the support of the Canadian Space
Agency. The data used in this work were taken as part of
Program ID M16AL002.

This research as made use of {\tt Astropy}, a community-developed core Python package for Astronomy (\citealt{Astropy2013, Astropy2018}). Physical parameters in IllustrisTNG were calculated using {\sc Colossus}, a public, open-source python package for calculations related to cosmology, the large-scale structure of matter in the universe, and the properties of dark matter halos \citep{Diemer2018_colossus}. This research also used the Python libraries {\tt SciPy} (\citealt{Scipy2020}), {\tt NumPy} (\citealt{Numpy}), {\tt Matplotlib} (\citealt{Matplotlib}) and {\tt pandas} (\citealt{pandas}). We thank Dr Gao, Siyu for the development of {\tt curlyBrace}.

\section*{Data availability}
The data used in this study is largely publicly available and can be shared upon reasonable request to the corresponding author.




\bibliographystyle{mnras}
\bibliography{TNGs2cosmos.bib} 

\begin{thebibliography}{}
\makeatletter
\relax
\def\mn@urlcharsother{\let\do\@makeother \do\$\do\&\do\#\do\^\do\_\do\%\do\~}
\def\mn@doi{\begingroup\mn@urlcharsother \@ifnextchar [ {\mn@doi@}
  {\mn@doi@[]}}
\def\mn@doi@[#1]#2{\def\@tempa{#1}\ifx\@tempa\@empty \href
  {http://dx.doi.org/#2} {doi:#2}\else \href {http://dx.doi.org/#2} {#1}\fi
  \endgroup}
\def\mn@eprint#1#2{\mn@eprint@#1:#2::\@nil}
\def\mn@eprint@arXiv#1{\href {http://arxiv.org/abs/#1} {{\tt arXiv:#1}}}
\def\mn@eprint@dblp#1{\href {http://dblp.uni-trier.de/rec/bibtex/#1.xml}
  {dblp:#1}}
\def\mn@eprint@#1:#2:#3:#4\@nil{\def\@tempa {#1}\def\@tempb {#2}\def\@tempc
  {#3}\ifx \@tempc \@empty \let \@tempc \@tempb \let \@tempb \@tempa \fi \ifx
  \@tempb \@empty \def\@tempb {arXiv}\fi \@ifundefined
  {mn@eprint@\@tempb}{\@tempb:\@tempc}{\expandafter \expandafter \csname
  mn@eprint@\@tempb\endcsname \expandafter{\@tempc}}}

\bibitem[\protect\citeauthoryear{{Abazajian} et~al.,}{{Abazajian}
  et~al.}{2009}]{Abazajian2009}
{Abazajian} K.~N.,  et~al., 2009, \mn@doi [\apjs]
  {10.1088/0067-0049/182/2/543}, \href
  {https://ui.adsabs.harvard.edu/abs/2009ApJS..182..543A} {182, 543}

\bibitem[\protect\citeauthoryear{{Ahn} et~al.,}{{Ahn} et~al.}{2014}]{Ahn2014}
{Ahn} C.~P.,  et~al., 2014, \mn@doi [The Astrophysical Journal Supplement
  Series] {10.1088/0067-0049/211/2/17}, \href
  {https://ui.adsabs.harvard.edu/\#abs/2014ApJS..211...17A} {211, 17}

\bibitem[\protect\citeauthoryear{{Andrews}, {Driver}, {Davies}, {Kafle},
  {Robotham}  \& {Wright}}{{Andrews} et~al.}{2017}]{Andrews2017}
{Andrews} S.~K.,  {Driver} S.~P.,  {Davies} L.~J.~M.,  {Kafle} P.~R.,
  {Robotham} A. S.~G.,   {Wright} A.~H.,  2017, \mn@doi [\mnras]
  {10.1093/mnras/stw2395}, \href
  {https://ui.adsabs.harvard.edu/\#abs/2017MNRAS.464.1569A} {464, 1569}

\bibitem[\protect\citeauthoryear{{Asano}, {Takeuchi}, {Hirashita}  \&
  {Inoue}}{{Asano} et~al.}{2013}]{Asano2013}
{Asano} R.~S.,  {Takeuchi} T.~T.,  {Hirashita} H.,   {Inoue} A.~K.,  2013,
  \mn@doi [Earth, Planets, and Space] {10.5047/eps.2012.04.014}, \href
  {https://ui.adsabs.harvard.edu/abs/2013EP&S...65..213A} {65, 213}

\bibitem[\protect\citeauthoryear{{Astropy Collaboration} et~al.,}{{Astropy
  Collaboration} et~al.}{2013}]{Astropy2013}
{Astropy Collaboration} et~al., 2013, \mn@doi [\aap]
  {10.1051/0004-6361/201322068}, \href
  {http://adsabs.harvard.edu/abs/2013A%26A...558A..33A} {558, A33}

\bibitem[\protect\citeauthoryear{{Baes}, {Verstappen}, {De Looze}, {Fritz},
  {Saftly}, {Vidal P{\'e}rez}, {Stalevski}  \& {Valcke}}{{Baes}
  et~al.}{2011}]{Baes2011}
{Baes} M.,  {Verstappen} J.,  {De Looze} I.,  {Fritz} J.,  {Saftly} W.,  {Vidal
  P{\'e}rez} E.,  {Stalevski} M.,   {Valcke} S.,  2011, \mn@doi [\apjs]
  {10.1088/0067-0049/196/2/22}, \href
  {https://ui.adsabs.harvard.edu/abs/2011ApJS..196...22B} {196, 22}

\bibitem[\protect\citeauthoryear{{Baes} et~al.,}{{Baes}
  et~al.}{2020}]{Baes2020}
{Baes} M.,  et~al., 2020, \mn@doi [\mnras] {10.1093/mnras/staa990}, \href
  {https://ui.adsabs.harvard.edu/abs/2020MNRAS.494.2912B} {494, 2912}

\bibitem[\protect\citeauthoryear{{Beeston} et~al.,}{{Beeston}
  et~al.}{2018}]{Beeston2018}
{Beeston} R.~A.,  et~al., 2018, \mn@doi [\mnras] {10.1093/mnras/sty1460}, \href
  {https://ui.adsabs.harvard.edu/abs/2018MNRAS.479.1077B} {479, 1077}

\bibitem[\protect\citeauthoryear{{B{\'e}thermin} et~al.,}{{B{\'e}thermin}
  et~al.}{2015}]{Bethermin2015}
{B{\'e}thermin} M.,  et~al., 2015, \mn@doi [\aap]
  {10.1051/0004-6361/201425031}, \href
  {https://ui.adsabs.harvard.edu/abs/2015A&A...573A.113B} {573, A113}

\bibitem[\protect\citeauthoryear{{Bolatto}, {Wolfire}  \& {Leroy}}{{Bolatto}
  et~al.}{2013}]{Bolatto2013}
{Bolatto} A.~D.,  {Wolfire} M.,   {Leroy} A.~K.,  2013, \mn@doi [\araa]
  {10.1146/annurev-astro-082812-140944}, \href
  {https://ui.adsabs.harvard.edu/abs/2013ARA&A..51..207B} {51, 207}

\bibitem[\protect\citeauthoryear{{Boselli} et~al.,}{{Boselli}
  et~al.}{2010}]{Boselli2010}
{Boselli} A.,  et~al., 2010, \mn@doi [\pasp] {10.1086/651535}, \href
  {https://ui.adsabs.harvard.edu/abs/2010PASP..122..261B} {122, 261}

\bibitem[\protect\citeauthoryear{{Bothwell} et~al.,}{{Bothwell}
  et~al.}{2013}]{Bothwell2013}
{Bothwell} M.~S.,  et~al., 2013, \mn@doi [\mnras] {10.1093/mnras/sts562}, \href
  {https://ui.adsabs.harvard.edu/abs/2013MNRAS.429.3047B} {429, 3047}

\bibitem[\protect\citeauthoryear{{Bruzual} \& {Charlot}}{{Bruzual} \&
  {Charlot}}{2003}]{BruzualCharlot2003}
{Bruzual} G.,  {Charlot} S.,  2003, \mn@doi [\mnras]
  {10.1046/j.1365-8711.2003.06897.x}, \href
  {http://adsabs.harvard.edu/abs/2003MNRAS.344.1000B} {344, 1000}

\bibitem[\protect\citeauthoryear{{Camps} \& {Baes}}{{Camps} \&
  {Baes}}{2015}]{CampsBaes2015}
{Camps} P.,  {Baes} M.,  2015, \mn@doi [Astronomy and Computing]
  {10.1016/j.ascom.2014.10.004}, \href
  {https://ui.adsabs.harvard.edu/abs/2015A&C.....9...20C} {9, 20}

\bibitem[\protect\citeauthoryear{{Camps}, {Trayford}, {Baes}, {Theuns},
  {Schaller}  \& {Schaye}}{{Camps} et~al.}{2016}]{Camps2016}
{Camps} P.,  {Trayford} J.~W.,  {Baes} M.,  {Theuns} T.,  {Schaller} M.,
  {Schaye} J.,  2016, \mn@doi [\mnras] {10.1093/mnras/stw1735}, \href
  {https://ui.adsabs.harvard.edu/abs/2016MNRAS.462.1057C} {462, 1057}

\bibitem[\protect\citeauthoryear{{Camps} et~al.,}{{Camps}
  et~al.}{2018}]{Camps2018}
{Camps} P.,  et~al., 2018, \mn@doi [\apjs] {10.3847/1538-4365/aaa24c}, \href
  {https://ui.adsabs.harvard.edu/abs/2018ApJS..234...20C} {234, 20}

\bibitem[\protect\citeauthoryear{{Carilli} \& {Walter}}{{Carilli} \&
  {Walter}}{2013}]{Carilli2013}
{Carilli} C.~L.,  {Walter} F.,  2013, \mn@doi [\araa]
  {10.1146/annurev-astro-082812-140953}, \href
  {https://ui.adsabs.harvard.edu/abs/2013ARA&A..51..105C} {51, 105}

\bibitem[\protect\citeauthoryear{{Casey} et~al.,}{{Casey}
  et~al.}{2011}]{Casey2011}
{Casey} C.~M.,  et~al., 2011, \mn@doi [\mnras]
  {10.1111/j.1365-2966.2011.18885.x}, \href
  {https://ui.adsabs.harvard.edu/abs/2011MNRAS.415.2723C} {415, 2723}

\bibitem[\protect\citeauthoryear{{Catinella} et~al.,}{{Catinella}
  et~al.}{2010}]{Catinella2010}
{Catinella} B.,  et~al., 2010, \mn@doi [\mnras]
  {10.1111/j.1365-2966.2009.16180.x}, \href
  {https://ui.adsabs.harvard.edu/abs/2010MNRAS.403..683C} {403, 683}

\bibitem[\protect\citeauthoryear{{Chabrier}}{{Chabrier}}{2003}]{Chabrier2003}
{Chabrier} G.,  2003, \mn@doi [\pasp] {10.1086/376392}, \href
  {http://adsabs.harvard.edu/abs/2003PASP..115..763C} {115, 763}

\bibitem[\protect\citeauthoryear{{Chapin} et~al.,}{{Chapin}
  et~al.}{2009}]{Chapin2009}
{Chapin} E.~L.,  et~al., 2009, \mn@doi [\mnras]
  {10.1111/j.1365-2966.2009.15267.x}, \href
  {https://ui.adsabs.harvard.edu/abs/2009MNRAS.398.1793C} {398, 1793}

\bibitem[\protect\citeauthoryear{{Charlot} \& {Fall}}{{Charlot} \&
  {Fall}}{2000}]{CharlotFall2000}
{Charlot} S.,  {Fall} S.~M.,  2000, \mn@doi [\apj] {10.1086/309250}, \href
  {http://adsabs.harvard.edu/abs/2000ApJ...539..718C} {539, 718}

\bibitem[\protect\citeauthoryear{{Chawner} et~al.,}{{Chawner}
  et~al.}{2019}]{Chawner2019}
{Chawner} H.,  et~al., 2019, \mn@doi [\mnras] {10.1093/mnras/sty2942}, \href
  {https://ui.adsabs.harvard.edu/abs/2019MNRAS.483...70C} {483, 70}

\bibitem[\protect\citeauthoryear{{Chawner} et~al.,}{{Chawner}
  et~al.}{2020}]{Chawner2020}
{Chawner} H.,  et~al., 2020, \mn@doi [\mnras] {10.1093/mnras/staa221}, \href
  {https://ui.adsabs.harvard.edu/abs/2020MNRAS.493.2706C} {493, 2706}

\bibitem[\protect\citeauthoryear{{Ciesla} et~al.,}{{Ciesla}
  et~al.}{2015}]{Ciesla2015}
{Ciesla} L.,  et~al., 2015, \mn@doi [\aap] {10.1051/0004-6361/201425252}, \href
  {https://ui.adsabs.harvard.edu/abs/2015A&A...576A..10C} {576, A10}

\bibitem[\protect\citeauthoryear{{Cigan} et~al.,}{{Cigan}
  et~al.}{2019}]{Cigan2019}
{Cigan} P.,  et~al., 2019, \mn@doi [\apj] {10.3847/1538-4357/ab4b46}, \href
  {https://ui.adsabs.harvard.edu/abs/2019ApJ...886...51C} {886, 51}

\bibitem[\protect\citeauthoryear{{Clark}, {Schofield}, {Gomez}  \&
  {Davies}}{{Clark} et~al.}{2016}]{Clark2016}
{Clark} C.~J.~R.,  {Schofield} S.~P.,  {Gomez} H.~L.,   {Davies} J.~I.,  2016,
  \mn@doi [\mnras] {10.1093/mnras/stw647}, \href
  {http://cdsads.u-strasbg.fr/abs/2016MNRAS.459.1646C} {459, 1646}

\bibitem[\protect\citeauthoryear{{Clemens} et~al.,}{{Clemens}
  et~al.}{2013}]{Clemens2013}
{Clemens} M.~S.,  et~al., 2013, \mn@doi [\mnras] {10.1093/mnras/stt760}, \href
  {https://ui.adsabs.harvard.edu/abs/2013MNRAS.433..695C} {433, 695}

\bibitem[\protect\citeauthoryear{{Clements}, {Dunne}  \& {Eales}}{{Clements}
  et~al.}{2010}]{Clements2010}
{Clements} D.~L.,  {Dunne} L.,   {Eales} S.,  2010, \mn@doi [\mnras]
  {10.1111/j.1365-2966.2009.16064.x}, \href
  {https://ui.adsabs.harvard.edu/abs/2010MNRAS.403..274C} {403, 274}

\bibitem[\protect\citeauthoryear{{Combes}}{{Combes}}{2018}]{Combes2018}
{Combes} F.,  2018, \mn@doi [\aapr] {10.1007/s00159-018-0110-4}, \href
  {https://ui.adsabs.harvard.edu/abs/2018A&ARv..26....5C} {26, 5}

\bibitem[\protect\citeauthoryear{{Cool} et~al.,}{{Cool}
  et~al.}{2013}]{Cool2013}
{Cool} R.~J.,  et~al., 2013, \mn@doi [\apj] {10.1088/0004-637X/767/2/118},
  \href {https://ui.adsabs.harvard.edu/\#abs/2013ApJ...767..118C} {767, 118}

\bibitem[\protect\citeauthoryear{{Coppin} et~al.,}{{Coppin}
  et~al.}{2009}]{Coppin2009}
{Coppin} K.~E.~K.,  et~al., 2009, \mn@doi [\mnras]
  {10.1111/j.1365-2966.2009.14700.x}, \href
  {https://ui.adsabs.harvard.edu/abs/2009MNRAS.395.1905C} {395, 1905}

\bibitem[\protect\citeauthoryear{{Crain} et~al.,}{{Crain}
  et~al.}{2015}]{Crain2015}
{Crain} R.~A.,  et~al., 2015, \mn@doi [\mnras] {10.1093/mnras/stv725}, \href
  {https://ui.adsabs.harvard.edu/abs/2015MNRAS.450.1937C} {450, 1937}

\bibitem[\protect\citeauthoryear{{Daddi} et~al.,}{{Daddi}
  et~al.}{2007}]{Daddi2007}
{Daddi} E.,  et~al., 2007, \mn@doi [\apj] {10.1086/521818}, \href
  {https://ui.adsabs.harvard.edu/abs/2007ApJ...670..156D} {670, 156}

\bibitem[\protect\citeauthoryear{{Dav{\'e}}, {Angl{\'e}s-Alc{\'a}zar},
  {Narayanan}, {Li}, {Rafieferantsoa}  \& {Appleby}}{{Dav{\'e}}
  et~al.}{2019}]{Dave2019}
{Dav{\'e}} R.,  {Angl{\'e}s-Alc{\'a}zar} D.,  {Narayanan} D.,  {Li} Q.,
  {Rafieferantsoa} M.~H.,   {Appleby} S.,  2019, \mn@doi [\mnras]
  {10.1093/mnras/stz937}, \href
  {https://ui.adsabs.harvard.edu/abs/2019MNRAS.486.2827D} {486, 2827}

\bibitem[\protect\citeauthoryear{{Dav{\'e}}, {Crain}, {Stevens}, {Narayanan},
  {Saintonge}, {Catinella}  \& {Cortese}}{{Dav{\'e}} et~al.}{2020}]{Dave2020}
{Dav{\'e}} R.,  {Crain} R.~A.,  {Stevens} A. R.~H.,  {Narayanan} D.,
  {Saintonge} A.,  {Catinella} B.,   {Cortese} L.,  2020, arXiv e-prints, \href
  {https://ui.adsabs.harvard.edu/abs/2020arXiv200207226D} {p. arXiv:2002.07226}

\bibitem[\protect\citeauthoryear{{Davies} et~al.,}{{Davies}
  et~al.}{2015}]{Davies2015}
{Davies} L.~J.~M.,  et~al., 2015, \mn@doi [\mnras] {10.1093/mnras/stu2515},
  \href {https://ui.adsabs.harvard.edu/\#abs/2015MNRAS.447.1014D} {447, 1014}

\bibitem[\protect\citeauthoryear{{Davis}, {Efstathiou}, {Frenk}  \&
  {White}}{{Davis} et~al.}{1985}]{Davis1985}
{Davis} M.,  {Efstathiou} G.,  {Frenk} C.~S.,   {White} S.~D.~M.,  1985,
  \mn@doi [\apj] {10.1086/163168}, \href
  {https://ui.adsabs.harvard.edu/abs/1985ApJ...292..371D} {292, 371}

\bibitem[\protect\citeauthoryear{{De Cia}, {Ledoux}, {Savaglio}, {Schady}  \&
  {Vreeswijk}}{{De Cia} et~al.}{2013}]{DeCia2013}
{De Cia} A.,  {Ledoux} C.,  {Savaglio} S.,  {Schady} P.,   {Vreeswijk} P.~M.,
  2013, \mn@doi [\aap] {10.1051/0004-6361/201321834}, \href
  {https://ui.adsabs.harvard.edu/abs/2013A&A...560A..88D} {560, A88}

\bibitem[\protect\citeauthoryear{{De Looze}, {Barlow}, {Swinyard}, {Rho},
  {Gomez}, {Matsuura}  \& {Wesson}}{{De Looze} et~al.}{2017}]{DeLooze2017}
{De Looze} I.,  {Barlow} M.~J.,  {Swinyard} B.~M.,  {Rho} J.,  {Gomez} H.~L.,
  {Matsuura} M.,   {Wesson} R.,  2017, \mn@doi [\mnras]
  {10.1093/mnras/stw2837}, \href
  {https://ui.adsabs.harvard.edu/abs/2017MNRAS.465.3309D} {465, 3309}

\bibitem[\protect\citeauthoryear{{De Looze} et~al.,}{{De Looze}
  et~al.}{2019}]{DeLooze2019}
{De Looze} I.,  et~al., 2019, \mn@doi [\mnras] {10.1093/mnras/stz1533}, \href
  {https://ui.adsabs.harvard.edu/abs/2019MNRAS.488..164D} {488, 164}

\bibitem[\protect\citeauthoryear{{De Vis} et~al.,}{{De Vis}
  et~al.}{2017a}]{deVis2017a}
{De Vis} P.,  et~al., 2017a, \mn@doi [\mnras] {10.1093/mnras/stw2501}, \href
  {https://ui.adsabs.harvard.edu/abs/2017MNRAS.464.4680D} {464, 4680}

\bibitem[\protect\citeauthoryear{{De Vis} et~al.,}{{De Vis}
  et~al.}{2017b}]{deVis2017b}
{De Vis} P.,  et~al., 2017b, \mn@doi [\mnras] {10.1093/mnras/stx981}, \href
  {https://ui.adsabs.harvard.edu/abs/2017MNRAS.471.1743D} {471, 1743}

\bibitem[\protect\citeauthoryear{{De Vis} et~al.,}{{De Vis}
  et~al.}{2019}]{devis2019}
{De Vis} P.,  et~al., 2019, \mn@doi [\aap] {10.1051/0004-6361/201834444}, \href
  {https://ui.adsabs.harvard.edu/abs/2019A&A...623A...5D} {623, A5}

\bibitem[\protect\citeauthoryear{Diemer}{Diemer}{2018}]{Diemer2018_colossus}
Diemer B.,  2018, \mn@doi [The Astrophysical Journal Supplement Series]
  {10.3847/1538-4365/aaee8c}, 239, 35

\bibitem[\protect\citeauthoryear{{Diemer} et~al.,}{{Diemer}
  et~al.}{2018}]{Diemer2018}
{Diemer} B.,  et~al., 2018, \mn@doi [\apjs] {10.3847/1538-4365/aae387}, \href
  {https://ui.adsabs.harvard.edu/abs/2018ApJS..238...33D} {238, 33}

\bibitem[\protect\citeauthoryear{{Diemer} et~al.,}{{Diemer}
  et~al.}{2019}]{Diemer2019}
{Diemer} B.,  et~al., 2019, \mn@doi [\mnras] {10.1093/mnras/stz1323}, \href
  {https://ui.adsabs.harvard.edu/abs/2019MNRAS.487.1529D} {487, 1529}

\bibitem[\protect\citeauthoryear{{Doherty}, {Gil-Pons}, {Lau}, {Lattanzio}  \&
  {Siess}}{{Doherty} et~al.}{2014}]{Doherty2014}
{Doherty} C.~L.,  {Gil-Pons} P.,  {Lau} H. H.~B.,  {Lattanzio} J.~C.,   {Siess}
  L.,  2014, \mn@doi [\mnras] {10.1093/mnras/stt1877}, \href
  {https://ui.adsabs.harvard.edu/abs/2014MNRAS.437..195D} {437, 195}

\bibitem[\protect\citeauthoryear{{Dolag}, {Borgani}, {Murante}  \&
  {Springel}}{{Dolag} et~al.}{2009}]{Dolag2009}
{Dolag} K.,  {Borgani} S.,  {Murante} G.,   {Springel} V.,  2009, \mn@doi
  [\mnras] {10.1111/j.1365-2966.2009.15034.x}, \href
  {https://ui.adsabs.harvard.edu/abs/2009MNRAS.399..497D} {399, 497}

\bibitem[\protect\citeauthoryear{{Dole} et~al.,}{{Dole}
  et~al.}{2006}]{Dole2006}
{Dole} H.,  et~al., 2006, \mn@doi [\aap] {10.1051/0004-6361:20054446}, \href
  {https://ui.adsabs.harvard.edu/abs/2006A&A...451..417D} {451, 417}

\bibitem[\protect\citeauthoryear{{Draine}}{{Draine}}{2009}]{Draine2009}
{Draine} B.~T.,  2009, in {Henning} T.,  {Gr{\"u}n} E.,   {Steinacker} J.,
  eds,  Astronomical Society of the Pacific Conference Series Vol. 414, Cosmic
  Dust - Near and Far. p.~453 (\mn@eprint {arXiv} {0903.1658})

\bibitem[\protect\citeauthoryear{{Driver} et~al.,}{{Driver}
  et~al.}{2011}]{Driver2011}
{Driver} S.~P.,  et~al., 2011, \mn@doi [\mnras]
  {10.1111/j.1365-2966.2010.18188.x}, \href
  {https://ui.adsabs.harvard.edu/abs/2011MNRAS.413..971D} {413, 971}

\bibitem[\protect\citeauthoryear{{Driver} et~al.,}{{Driver}
  et~al.}{2018}]{Driver2018}
{Driver} S.~P.,  et~al., 2018, \mn@doi [\mnras] {10.1093/mnras/stx2728}, \href
  {https://ui.adsabs.harvard.edu/\#abs/2018MNRAS.475.2891D} {475, 2891}

\bibitem[\protect\citeauthoryear{{Dunne} \& {Eales}}{{Dunne} \&
  {Eales}}{2001}]{Dunne2001}
{Dunne} L.,  {Eales} S.~A.,  2001, \mn@doi [\mnras]
  {10.1046/j.1365-8711.2001.04789.x}, \href
  {https://ui.adsabs.harvard.edu/abs/2001MNRAS.327..697D} {327, 697}

\bibitem[\protect\citeauthoryear{{Dunne}, {Eales}, {Edmunds}, {Ivison},
  {Alexander}  \& {Clements}}{{Dunne} et~al.}{2000}]{Dunne2000}
{Dunne} L.,  {Eales} S.,  {Edmunds} M.,  {Ivison} R.,  {Alexander} P.,
  {Clements} D.~L.,  2000, \mn@doi [\mnras] {10.1046/j.1365-8711.2000.03386.x},
  \href {https://ui.adsabs.harvard.edu/abs/2000MNRAS.315..115D} {315, 115}

\bibitem[\protect\citeauthoryear{{Dunne}, {Eales}  \& {Edmunds}}{{Dunne}
  et~al.}{2003a}]{Dunne2003}
{Dunne} L.,  {Eales} S.~A.,   {Edmunds} M.~G.,  2003a, \mn@doi [\mnras]
  {10.1046/j.1365-8711.2003.06440.x}, \href
  {https://ui.adsabs.harvard.edu/abs/2003MNRAS.341..589D} {341, 589}

\bibitem[\protect\citeauthoryear{{Dunne}, {Eales}, {Ivison}, {Morgan}  \&
  {Edmunds}}{{Dunne} et~al.}{2003b}]{Dunne2003SneDust}
{Dunne} L.,  {Eales} S.,  {Ivison} R.,  {Morgan} H.,   {Edmunds} M.,  2003b,
  \mn@doi [\nat] {10.1038/nature01792}, \href
  {https://ui.adsabs.harvard.edu/abs/2003Natur.424..285D} {424, 285}

\bibitem[\protect\citeauthoryear{{Dunne} et~al.,}{{Dunne}
  et~al.}{2011}]{Dunne2011}
{Dunne} L.,  et~al., 2011, \mn@doi [\mnras] {10.1111/j.1365-2966.2011.19363.x},
  \href {https://ui.adsabs.harvard.edu/abs/2011MNRAS.417.1510D} {417, 1510}

\bibitem[\protect\citeauthoryear{{Dwek}, {Galliano}  \& {Jones}}{{Dwek}
  et~al.}{2007}]{Dwek2007}
{Dwek} E.,  {Galliano} F.,   {Jones} A.~P.,  2007, \mn@doi [\apj]
  {10.1086/518430}, \href
  {https://ui.adsabs.harvard.edu/abs/2007ApJ...662..927D} {662, 927}

\bibitem[\protect\citeauthoryear{{Eales} et~al.,}{{Eales}
  et~al.}{2009}]{Eales2009}
{Eales} S.,  et~al., 2009, \mn@doi [\apj] {10.1088/0004-637X/707/2/1779}, \href
  {https://ui.adsabs.harvard.edu/abs/2009ApJ...707.1779E} {707, 1779}

\bibitem[\protect\citeauthoryear{{Eales} et~al.,}{{Eales}
  et~al.}{2010}]{Eales2010}
{Eales} S.,  et~al., 2010, \mn@doi [\pasp] {10.1086/653086}, \href
  {https://ui.adsabs.harvard.edu/abs/2010PASP..122..499E} {122, 499}

\bibitem[\protect\citeauthoryear{{Eales} et~al.,}{{Eales}
  et~al.}{2012}]{Eales2012}
{Eales} S.,  et~al., 2012, \mn@doi [\apj] {10.1088/0004-637X/761/2/168}, \href
  {http://adsabs.harvard.edu/abs/2012ApJ...761..168E} {761, 168}

\bibitem[\protect\citeauthoryear{{Ferland}, {Korista}, {Verner}, {Ferguson},
  {Kingdon}  \& {Verner}}{{Ferland} et~al.}{1998}]{Ferland1998}
{Ferland} G.~J.,  {Korista} K.~T.,  {Verner} D.~A.,  {Ferguson} J.~W.,
  {Kingdon} J.~B.,   {Verner} E.~M.,  1998, \mn@doi [\pasp] {10.1086/316190},
  \href {https://ui.adsabs.harvard.edu/abs/1998PASP..110..761F} {110, 761}

\bibitem[\protect\citeauthoryear{{Fishlock}, {Karakas}, {Lugaro}  \&
  {Yong}}{{Fishlock} et~al.}{2014}]{Fishlock2014}
{Fishlock} C.~K.,  {Karakas} A.~I.,  {Lugaro} M.,   {Yong} D.,  2014, \mn@doi
  [\apj] {10.1088/0004-637X/797/1/44}, \href
  {https://ui.adsabs.harvard.edu/abs/2014ApJ...797...44F} {797, 44}

\bibitem[\protect\citeauthoryear{{Fixsen}, {Dwek}, {Mather}, {Bennett}  \&
  {Shafer}}{{Fixsen} et~al.}{1998}]{Fixsen1998}
{Fixsen} D.~J.,  {Dwek} E.,  {Mather} J.~C.,  {Bennett} C.~L.,   {Shafer}
  R.~A.,  1998, \mn@doi [\apj] {10.1086/306383}, \href
  {https://ui.adsabs.harvard.edu/abs/1998ApJ...508..123F} {508, 123}

\bibitem[\protect\citeauthoryear{{Geach} et~al.,}{{Geach}
  et~al.}{2017}]{Geach2017}
{Geach} J.~E.,  et~al., 2017, \mn@doi [\mnras] {10.1093/mnras/stw2721}, \href
  {https://ui.adsabs.harvard.edu/abs/2017MNRAS.465.1789G} {465, 1789}

\bibitem[\protect\citeauthoryear{{Genel} et~al.,}{{Genel}
  et~al.}{2014}]{Genel2014}
{Genel} S.,  et~al., 2014, \mn@doi [\mnras] {10.1093/mnras/stu1654}, \href
  {https://ui.adsabs.harvard.edu/abs/2014MNRAS.445..175G} {445, 175}

\bibitem[\protect\citeauthoryear{{Genzel} et~al.,}{{Genzel}
  et~al.}{2015}]{Genzel2015}
{Genzel} R.,  et~al., 2015, \mn@doi [\apj] {10.1088/0004-637X/800/1/20}, \href
  {http://adsabs.harvard.edu/abs/2015ApJ...800...20G} {800, 20}

\bibitem[\protect\citeauthoryear{{Gomez} et~al.,}{{Gomez}
  et~al.}{2012}]{Gomez2012}
{Gomez} H.~L.,  et~al., 2012, \mn@doi [\apj] {10.1088/0004-637X/760/1/96},
  \href {https://ui.adsabs.harvard.edu/abs/2012ApJ...760...96G} {760, 96}

\bibitem[\protect\citeauthoryear{Hayward et~al.,}{Hayward
  et~al.}{2020}]{Hayward2020}
Hayward C.~C.,  et~al., 2020, Submillimetre galaxies in cosmological
  hydrodynamical simulations -- an opportunity for constraining feedback models
  (\mn@eprint {arXiv} {2007.01885})

\bibitem[\protect\citeauthoryear{{Hildebrand}}{{Hildebrand}}{1983}]{Hildebrand1983}
{Hildebrand} R.~H.,  1983, \qjras, \href
  {http://adsabs.harvard.edu/abs/1983QJRAS..24..267H} {24, 267}

\bibitem[\protect\citeauthoryear{{Holland} et~al.,}{{Holland}
  et~al.}{2013}]{Holland2013}
{Holland} W.~S.,  et~al., 2013, \mn@doi [\mnras] {10.1093/mnras/sts612}, \href
  {https://ui.adsabs.harvard.edu/abs/2013MNRAS.430.2513H} {430, 2513}

\bibitem[\protect\citeauthoryear{{Hunter}}{{Hunter}}{2007}]{Matplotlib}
{Hunter} J.~D.,  2007, Computing in Science Engineering, 9, 90

\bibitem[\protect\citeauthoryear{{Inoue}}{{Inoue}}{2003}]{Inoue2003}
{Inoue} A.~K.,  2003, \mn@doi [\pasj] {10.1093/pasj/55.5.901}, \href
  {https://ui.adsabs.harvard.edu/abs/2003PASJ...55..901I} {55, 901}

\bibitem[\protect\citeauthoryear{{James}, {Dunne}, {Eales}  \&
  {Edmunds}}{{James} et~al.}{2002}]{James2002}
{James} A.,  {Dunne} L.,  {Eales} S.,   {Edmunds} M.~G.,  2002, \mn@doi
  [\mnras] {10.1046/j.1365-8711.2002.05660.x}, \href
  {https://ui.adsabs.harvard.edu/abs/2002MNRAS.335..753J} {335, 753}

\bibitem[\protect\citeauthoryear{{Karakas}}{{Karakas}}{2010}]{Karakas2010}
{Karakas} A.~I.,  2010, \mn@doi [\mnras] {10.1111/j.1365-2966.2009.16198.x},
  \href {https://ui.adsabs.harvard.edu/abs/2010MNRAS.403.1413K} {403, 1413}

\bibitem[\protect\citeauthoryear{{Karim} et~al.,}{{Karim}
  et~al.}{2011}]{Karim2011}
{Karim} A.,  et~al., 2011, \mn@doi [\apj] {10.1088/0004-637X/730/2/61}, \href
  {https://ui.adsabs.harvard.edu/abs/2011ApJ...730...61K} {730, 61}

\bibitem[\protect\citeauthoryear{{Kobayashi}, {Umeda}, {Nomoto}, {Tominaga}  \&
  {Ohkubo}}{{Kobayashi} et~al.}{2006}]{Kobayashi2006}
{Kobayashi} C.,  {Umeda} H.,  {Nomoto} K.,  {Tominaga} N.,   {Ohkubo} T.,
  2006, \mn@doi [\apj] {10.1086/508914}, \href
  {https://ui.adsabs.harvard.edu/abs/2006ApJ...653.1145K} {653, 1145}

\bibitem[\protect\citeauthoryear{{Lagos} et~al.,}{{Lagos}
  et~al.}{2019}]{Lagos2019}
{Lagos} C. d.~P.,  et~al., 2019, \mn@doi [\mnras] {10.1093/mnras/stz2427},
  \href {https://ui.adsabs.harvard.edu/abs/2019MNRAS.489.4196L} {489, 4196}

\bibitem[\protect\citeauthoryear{{Laigle} et~al.,}{{Laigle}
  et~al.}{2016}]{Laigle2016}
{Laigle} C.,  et~al., 2016, \mn@doi [The Astrophysical Journal Supplement
  Series] {10.3847/0067-0049/224/2/24}, \href
  {https://ui.adsabs.harvard.edu/\#abs/2016ApJS..224...24L} {224, 24}

\bibitem[\protect\citeauthoryear{{Le F{\`e}vre} et~al.,}{{Le F{\`e}vre}
  et~al.}{2013}]{LeFevre2013}
{Le F{\`e}vre} O.,  et~al., 2013, \mn@doi [\aap] {10.1051/0004-6361/201322179},
  \href {https://ui.adsabs.harvard.edu/\#abs/2013A&A...559A..14L} {559, A14}

\bibitem[\protect\citeauthoryear{{Lee} et~al.,}{{Lee} et~al.}{2015}]{Lee2015}
{Lee} N.,  et~al., 2015, \mn@doi [\apj] {10.1088/0004-637X/801/2/80}, \href
  {https://ui.adsabs.harvard.edu/abs/2015ApJ...801...80L} {801, 80}

\bibitem[\protect\citeauthoryear{{Li}, {Narayanan}  \& {Dav{\'e}}}{{Li}
  et~al.}{2019}]{Li2019}
{Li} Q.,  {Narayanan} D.,   {Dav{\'e}} R.,  2019, \mn@doi [\mnras]
  {10.1093/mnras/stz2684}, \href
  {https://ui.adsabs.harvard.edu/abs/2019MNRAS.490.1425L} {490, 1425}

\bibitem[\protect\citeauthoryear{{Liang} et~al.,}{{Liang}
  et~al.}{2019}]{Liang2019}
{Liang} L.,  et~al., 2019, \mn@doi [\mnras] {10.1093/mnras/stz2134}, \href
  {https://ui.adsabs.harvard.edu/abs/2019MNRAS.tmp.2072L} {p.~2072}

\bibitem[\protect\citeauthoryear{{Lilly} et~al.,}{{Lilly}
  et~al.}{2007}]{Lilly2007zbright}
{Lilly} S.~J.,  et~al., 2007, \mn@doi [The Astrophysical Journal Supplement
  Series] {10.1086/516589}, \href
  {https://ui.adsabs.harvard.edu/\#abs/2007ApJS..172...70L} {172, 70}

\bibitem[\protect\citeauthoryear{{Madau} \& {Dickinson}}{{Madau} \&
  {Dickinson}}{2014}]{Madau2014}
{Madau} P.,  {Dickinson} M.,  2014, \mn@doi [Annual Review of Astronomy and
  Astrophysics] {10.1146/annurev-astro-081811-125615}, \href
  {https://ui.adsabs.harvard.edu/\#abs/2014ARA&A..52..415M} {52, 415}

\bibitem[\protect\citeauthoryear{{Madden} et~al.,}{{Madden}
  et~al.}{2013}]{Madden2013}
{Madden} S.~C.,  et~al., 2013, \mn@doi [\pasp] {10.1086/671138}, \href
  {https://ui.adsabs.harvard.edu/abs/2013PASP..125..600M} {125, 600}

\bibitem[\protect\citeauthoryear{{Marinacci} et~al.,}{{Marinacci}
  et~al.}{2018}]{Marinacci2018}
{Marinacci} F.,  et~al., 2018, \mn@doi [\mnras] {10.1093/mnras/sty2206}, \href
  {https://ui.adsabs.harvard.edu/abs/2018MNRAS.480.5113M} {480, 5113}

\bibitem[\protect\citeauthoryear{{Matsuura} et~al.,}{{Matsuura}
  et~al.}{2011}]{Matsuura2011}
{Matsuura} M.,  et~al., 2011, \mn@doi [Science] {10.1126/science.1205983},
  \href {https://ui.adsabs.harvard.edu/abs/2011Sci...333.1258M} {333, 1258}

\bibitem[\protect\citeauthoryear{McKinney}{McKinney}{2010}]{pandas}
McKinney W.,  2010, in van~der Walt S.,  Millman J.,  eds, Proceedings of the
  9th Python in Science Conference. pp 51 -- 56

\bibitem[\protect\citeauthoryear{{McKinnon}, {Torrey}  \&
  {Vogelsberger}}{{McKinnon} et~al.}{2016}]{McKinnon2016}
{McKinnon} R.,  {Torrey} P.,   {Vogelsberger} M.,  2016, \mn@doi [\mnras]
  {10.1093/mnras/stw253}, \href
  {https://ui.adsabs.harvard.edu/abs/2016MNRAS.457.3775M} {457, 3775}

\bibitem[\protect\citeauthoryear{{McKinnon}, {Kannan}, {Vogelsberger},
  {O'Neil}, {Torrey}  \& {Li}}{{McKinnon} et~al.}{2019}]{McKinnon2019}
{McKinnon} R.,  {Kannan} R.,  {Vogelsberger} M.,  {O'Neil} S.,  {Torrey} P.,
  {Li} H.,  2019, arXiv e-prints, \href
  {https://ui.adsabs.harvard.edu/abs/2019arXiv191202825M} {p. arXiv:1912.02825}

\bibitem[\protect\citeauthoryear{{Meyer}, {Jura}  \& {Cardelli}}{{Meyer}
  et~al.}{1998}]{Meyer1998}
{Meyer} D.~M.,  {Jura} M.,   {Cardelli} J.~A.,  1998, \mn@doi [\apj]
  {10.1086/305128}, \href
  {https://ui.adsabs.harvard.edu/abs/1998ApJ...493..222M} {493, 222}

\bibitem[\protect\citeauthoryear{{Micha{\l}owski} et~al.,}{{Micha{\l}owski}
  et~al.}{2017}]{Michalowski2017}
{Micha{\l}owski} M.~J.,  et~al., 2017, \mn@doi [\mnras] {10.1093/mnras/stx861},
  \href {https://ui.adsabs.harvard.edu/abs/2017MNRAS.469..492M} {469, 492}

\bibitem[\protect\citeauthoryear{{Millard} et~al.,}{{Millard}
  et~al.}{2020}]{Millard2020}
{Millard} J.~S.,  et~al., 2020, \mn@doi [\mnras] {10.1093/mnras/staa609}, \href
  {https://ui.adsabs.harvard.edu/abs/2020MNRAS.tmp..573M} {}

\bibitem[\protect\citeauthoryear{{Morgan} \& {Edmunds}}{{Morgan} \&
  {Edmunds}}{2003}]{Morgan2003}
{Morgan} H.~L.,  {Edmunds} M.~G.,  2003, \mn@doi [\mnras]
  {10.1046/j.1365-8711.2003.06681.x}, \href
  {https://ui.adsabs.harvard.edu/abs/2003MNRAS.343..427M} {343, 427}

\bibitem[\protect\citeauthoryear{{Naiman} et~al.,}{{Naiman}
  et~al.}{2018}]{Naiman2018}
{Naiman} J.~P.,  et~al., 2018, \mn@doi [\mnras] {10.1093/mnras/sty618}, \href
  {https://ui.adsabs.harvard.edu/abs/2018MNRAS.477.1206N} {477, 1206}

\bibitem[\protect\citeauthoryear{{Nelson} et~al.,}{{Nelson}
  et~al.}{2018a}]{Nelson2018}
{Nelson} D.,  et~al., 2018a, \mn@doi [\mnras] {10.1093/mnras/stx3040}, \href
  {https://ui.adsabs.harvard.edu/abs/2018MNRAS.475..624N} {475, 624}

\bibitem[\protect\citeauthoryear{Nelson et~al.,}{Nelson
  et~al.}{2018b}]{Nelson2018b}
Nelson D.,  et~al., 2018b, \mn@doi [Monthly Notices of the Royal Astronomical
  Society] {10.1093/mnras/sty656}, 477, 450

\bibitem[\protect\citeauthoryear{{Nelson} et~al.,}{{Nelson}
  et~al.}{2019}]{Nelson2019}
{Nelson} D.,  et~al., 2019, \mn@doi [Computational Astrophysics and Cosmology]
  {10.1186/s40668-019-0028-x}, \href
  {https://ui.adsabs.harvard.edu/abs/2019ComAC...6....2N} {6, 2}

\bibitem[\protect\citeauthoryear{{Nomoto}, {Iwamoto}, {Nakasato}, {Thielemann},
  {Brachwitz}, {Tsujimoto}, {Kubo}  \& {Kishimoto}}{{Nomoto}
  et~al.}{1997}]{Nomoto1997}
{Nomoto} K.,  {Iwamoto} K.,  {Nakasato} N.,  {Thielemann} F.~K.,  {Brachwitz}
  F.,  {Tsujimoto} T.,  {Kubo} Y.,   {Kishimoto} N.,  1997, \mn@doi [\nphysa]
  {10.1016/S0375-9474(97)00291-1}, \href
  {https://ui.adsabs.harvard.edu/abs/1997NuPhA.621..467N} {621, 467}

\bibitem[\protect\citeauthoryear{{Pilbratt} et~al.,}{{Pilbratt}
  et~al.}{2010}]{Pilbratt2010}
{Pilbratt} G.~L.,  et~al., 2010, \mn@doi [\aap] {10.1051/0004-6361/201014759},
  \href {https://ui.adsabs.harvard.edu/\#abs/2010A&A...518L...1P} {518, L1}

\bibitem[\protect\citeauthoryear{{Pillepich} et~al.,}{{Pillepich}
  et~al.}{2018a}]{Pillepich2018a}
{Pillepich} A.,  et~al., 2018a, \mn@doi [\mnras] {10.1093/mnras/stx2656}, \href
  {https://ui.adsabs.harvard.edu/abs/2018MNRAS.473.4077P} {473, 4077}

\bibitem[\protect\citeauthoryear{{Pillepich} et~al.,}{{Pillepich}
  et~al.}{2018b}]{Pillepich2018b}
{Pillepich} A.,  et~al., 2018b, \mn@doi [\mnras] {10.1093/mnras/stx3112}, \href
  {https://ui.adsabs.harvard.edu/abs/2018MNRAS.475..648P} {475, 648}

\bibitem[\protect\citeauthoryear{{Planck Collaboration} et~al.,}{{Planck
  Collaboration} et~al.}{2016}]{Planck2015}
{Planck Collaboration} et~al., 2016, \mn@doi [\aap]
  {10.1051/0004-6361/201525830}, \href
  {https://ui.adsabs.harvard.edu/abs/2016A&A...594A..13P} {594, A13}

\bibitem[\protect\citeauthoryear{{Popping} et~al.,}{{Popping}
  et~al.}{2019}]{Popping2019}
{Popping} G.,  et~al., 2019, \mn@doi [\apj] {10.3847/1538-4357/ab30f2}, \href
  {https://ui.adsabs.harvard.edu/abs/2019ApJ...882..137P} {882, 137}

\bibitem[\protect\citeauthoryear{{Portinari}, {Chiosi}  \&
  {Bressan}}{{Portinari} et~al.}{1998}]{Portinari1998}
{Portinari} L.,  {Chiosi} C.,   {Bressan} A.,  1998, \aap, \href
  {https://ui.adsabs.harvard.edu/abs/1998A&A...334..505P} {334, 505}

\bibitem[\protect\citeauthoryear{{Price-Whelan} et~al.,}{{Price-Whelan}
  et~al.}{2018}]{Astropy2018}
{Price-Whelan} A.~M.,  et~al., 2018, \mn@doi [\aj] {10.3847/1538-3881/aabc4f},
  \href {https://ui.adsabs.harvard.edu/#abs/2018AJ....156..123T} {156, 123}

\bibitem[\protect\citeauthoryear{{Puget}, {Abergel}, {Bernard}, {Boulanger},
  {Burton}, {Desert}  \& {Hartmann}}{{Puget} et~al.}{1996}]{Puget1996}
{Puget} J.~L.,  {Abergel} A.,  {Bernard} J.~P.,  {Boulanger} F.,  {Burton}
  W.~B.,  {Desert} F.~X.,   {Hartmann} D.,  1996, \aap, \href
  {https://ui.adsabs.harvard.edu/abs/1996A&A...308L...5P} {308, L5}

\bibitem[\protect\citeauthoryear{{R{\'e}my-Ruyer} et~al.,}{{R{\'e}my-Ruyer}
  et~al.}{2014}]{Remy-Ruyer2014}
{R{\'e}my-Ruyer} A.,  et~al., 2014, \mn@doi [\aap]
  {10.1051/0004-6361/201322803}, \href
  {https://ui.adsabs.harvard.edu/abs/2014A&A...563A..31R} {563, A31}

\bibitem[\protect\citeauthoryear{{Roman-Duval} et~al.,}{{Roman-Duval}
  et~al.}{2019}]{RomanDuval2019}
{Roman-Duval} J.,  et~al., 2019, \mn@doi [\apj] {10.3847/1538-4357/aaf8bb},
  \href {https://ui.adsabs.harvard.edu/abs/2019ApJ...871..151R} {871, 151}

\bibitem[\protect\citeauthoryear{{Salpeter}}{{Salpeter}}{1974}]{Salpeter1974}
{Salpeter} E.~E.,  1974, \mn@doi [\apj] {10.1086/153195}, \href
  {https://ui.adsabs.harvard.edu/abs/1974ApJ...193..579S} {193, 579}

\bibitem[\protect\citeauthoryear{{Schaye} et~al.,}{{Schaye}
  et~al.}{2015}]{Schaye2015}
{Schaye} J.,  et~al., 2015, \mn@doi [\mnras] {10.1093/mnras/stu2058}, \href
  {https://ui.adsabs.harvard.edu/abs/2015MNRAS.446..521S} {446, 521}

\bibitem[\protect\citeauthoryear{{Schulz}, {Popping}, {Pillepich}, {Nelson},
  {Vogelsberger}, {Marinacci}  \& {Hernquist}}{{Schulz}
  et~al.}{2020}]{Schulz2020}
{Schulz} S.,  {Popping} G.,  {Pillepich} A.,  {Nelson} D.,  {Vogelsberger} M.,
  {Marinacci} F.,   {Hernquist} L.,  2020, arXiv e-prints, \href
  {https://ui.adsabs.harvard.edu/abs/2020arXiv200104992S} {p. arXiv:2001.04992}

\bibitem[\protect\citeauthoryear{{Scoville} et~al.,}{{Scoville}
  et~al.}{2007}]{Scoville2007}
{Scoville} N.,  et~al., 2007, \mn@doi [The Astrophysical Journal Supplement
  Series] {10.1086/516585}, \href
  {https://ui.adsabs.harvard.edu/abs/2007ApJS..172....1S} {172, 1}

\bibitem[\protect\citeauthoryear{{Scoville} et~al.,}{{Scoville}
  et~al.}{2014}]{Scoville2014}
{Scoville} N.,  et~al., 2014, \mn@doi [\apj] {10.1088/0004-637X/783/2/84},
  \href {https://ui.adsabs.harvard.edu/abs/2014ApJ...783...84S} {783, 84}

\bibitem[\protect\citeauthoryear{{Scoville} et~al.,}{{Scoville}
  et~al.}{2016}]{Scoville2016}
{Scoville} N.,  et~al., 2016, \mn@doi [\apj] {10.3847/0004-637X/820/2/83},
  \href {http://adsabs.harvard.edu/abs/2016ApJ...820...83S} {820, 83}

\bibitem[\protect\citeauthoryear{{Scoville} et~al.,}{{Scoville}
  et~al.}{2017}]{Scoville2017}
{Scoville} N.,  et~al., 2017, \mn@doi [\apj] {10.3847/1538-4357/aa61a0}, \href
  {http://adsabs.harvard.edu/abs/2017ApJ...837..150S} {837, 150}

\bibitem[\protect\citeauthoryear{{Shen} et~al.,}{{Shen}
  et~al.}{2020}]{Shen2020}
{Shen} X.,  et~al., 2020, arXiv e-prints, \href
  {https://ui.adsabs.harvard.edu/abs/2020arXiv200210474S} {p. arXiv:2002.10474}

\bibitem[\protect\citeauthoryear{{Sijacki}, {Vogelsberger}, {Genel},
  {Springel}, {Torrey}, {Snyder}, {Nelson}  \& {Hernquist}}{{Sijacki}
  et~al.}{2015}]{Sijacki2015}
{Sijacki} D.,  {Vogelsberger} M.,  {Genel} S.,  {Springel} V.,  {Torrey} P.,
  {Snyder} G.~F.,  {Nelson} D.,   {Hernquist} L.,  2015, \mn@doi [\mnras]
  {10.1093/mnras/stv1340}, \href
  {https://ui.adsabs.harvard.edu/abs/2015MNRAS.452..575S} {452, 575}

\bibitem[\protect\citeauthoryear{{Simpson} et~al.,}{{Simpson}
  et~al.}{2019}]{Simpson2019}
{Simpson} J.~M.,  et~al., 2019, \mn@doi [\apj] {10.3847/1538-4357/ab23ff},
  \href {https://ui.adsabs.harvard.edu/abs/2019ApJ...880...43S} {880, 43}

\bibitem[\protect\citeauthoryear{{Solomon} \& {Vanden Bout}}{{Solomon} \&
  {Vanden Bout}}{2005}]{Solomon2005}
{Solomon} P.~M.,  {Vanden Bout} P.~A.,  2005, \mn@doi [\araa]
  {10.1146/annurev.astro.43.051804.102221}, \href
  {https://ui.adsabs.harvard.edu/abs/2005ARA&A..43..677S} {43, 677}

\bibitem[\protect\citeauthoryear{{Springel}}{{Springel}}{2010}]{Springel2010}
{Springel} V.,  2010, \mn@doi [\mnras] {10.1111/j.1365-2966.2009.15715.x},
  \href {https://ui.adsabs.harvard.edu/abs/2010MNRAS.401..791S} {401, 791}

\bibitem[\protect\citeauthoryear{{Springel} \& {Hernquist}}{{Springel} \&
  {Hernquist}}{2003}]{SpringelHernquist2003}
{Springel} V.,  {Hernquist} L.,  2003, \mn@doi [\mnras]
  {10.1046/j.1365-8711.2003.06206.x}, \href
  {https://ui.adsabs.harvard.edu/abs/2003MNRAS.339..289S} {339, 289}

\bibitem[\protect\citeauthoryear{{Springel}, {White}, {Tormen}  \&
  {Kauffmann}}{{Springel} et~al.}{2001}]{Springel2001}
{Springel} V.,  {White} S. D.~M.,  {Tormen} G.,   {Kauffmann} G.,  2001,
  \mn@doi [\mnras] {10.1046/j.1365-8711.2001.04912.x}, \href
  {https://ui.adsabs.harvard.edu/abs/2001MNRAS.328..726S} {328, 726}

\bibitem[\protect\citeauthoryear{{Springel} et~al.,}{{Springel}
  et~al.}{2018}]{Springel2018}
{Springel} V.,  et~al., 2018, \mn@doi [\mnras] {10.1093/mnras/stx3304}, \href
  {https://ui.adsabs.harvard.edu/abs/2018MNRAS.475..676S} {475, 676}

\bibitem[\protect\citeauthoryear{{Stevens} et~al.,}{{Stevens}
  et~al.}{2019}]{Stevens2019}
{Stevens} A. R.~H.,  et~al., 2019, \mn@doi [\mnras] {10.1093/mnras/sty3451},
  \href {https://ui.adsabs.harvard.edu/abs/2019MNRAS.483.5334S} {483, 5334}

\bibitem[\protect\citeauthoryear{{Tacconi} et~al.,}{{Tacconi}
  et~al.}{2010}]{Tacconi2010}
{Tacconi} L.~J.,  et~al., 2010, \mn@doi [\nat] {10.1038/nature08773}, \href
  {https://ui.adsabs.harvard.edu/abs/2010Natur.463..781T} {463, 781}

\bibitem[\protect\citeauthoryear{{Tacconi} et~al.,}{{Tacconi}
  et~al.}{2013}]{Tacconi2013}
{Tacconi} L.~J.,  et~al., 2013, \mn@doi [\apj] {10.1088/0004-637X/768/1/74},
  \href {https://ui.adsabs.harvard.edu/abs/2013ApJ...768...74T} {768, 74}

\bibitem[\protect\citeauthoryear{{Tacconi} et~al.,}{{Tacconi}
  et~al.}{2018}]{Tacconi2018}
{Tacconi} L.~J.,  et~al., 2018, \mn@doi [\apj] {10.3847/1538-4357/aaa4b4},
  \href {http://adsabs.harvard.edu/abs/2018ApJ...853..179T} {853, 179}

\bibitem[\protect\citeauthoryear{{Terrazas} et~al.,}{{Terrazas}
  et~al.}{2020}]{Terrazas2020}
{Terrazas} B.~A.,  et~al., 2020, \mn@doi [\mnras] {10.1093/mnras/staa374},
  \href {https://ui.adsabs.harvard.edu/abs/2020MNRAS.493.1888T} {493, 1888}

\bibitem[\protect\citeauthoryear{Torrey, Vogelsberger, Sijacki, Springel  \&
  Hernquist}{Torrey et~al.}{2012}]{Torrey2012}
Torrey P.,  Vogelsberger M.,  Sijacki D.,  Springel V.,   Hernquist L.,  2012,
  \mn@doi [Monthly Notices of the Royal Astronomical Society]
  {10.1111/j.1365-2966.2012.22082.x}, 427, 2224–2238

\bibitem[\protect\citeauthoryear{{Torrey}, {Vogelsberger}, {Genel}, {Sijacki},
  {Springel}  \& {Hernquist}}{{Torrey} et~al.}{2014}]{Torrey2014}
{Torrey} P.,  {Vogelsberger} M.,  {Genel} S.,  {Sijacki} D.,  {Springel} V.,
  {Hernquist} L.,  2014, \mn@doi [\mnras] {10.1093/mnras/stt2295}, \href
  {https://ui.adsabs.harvard.edu/abs/2014MNRAS.438.1985T} {438, 1985}

\bibitem[\protect\citeauthoryear{{Torrey} et~al.,}{{Torrey}
  et~al.}{2019}]{Torrey2019}
{Torrey} P.,  et~al., 2019, \mn@doi [\mnras] {10.1093/mnras/stz243}, \href
  {https://ui.adsabs.harvard.edu/abs/2019MNRAS.484.5587T} {484, 5587}

\bibitem[\protect\citeauthoryear{{Valiante}, {Schneider}, {Bianchi}  \&
  {Andersen}}{{Valiante} et~al.}{2009}]{Valiante2009}
{Valiante} R.,  {Schneider} R.,  {Bianchi} S.,   {Andersen} A.~C.,  2009,
  \mn@doi [\mnras] {10.1111/j.1365-2966.2009.15076.x}, \href
  {https://ui.adsabs.harvard.edu/abs/2009MNRAS.397.1661V} {397, 1661}

\bibitem[\protect\citeauthoryear{{Virtanen} et~al.,}{{Virtanen}
  et~al.}{2020}]{Scipy2020}
{Virtanen} P.,  et~al., 2020, \mn@doi [Nature Methods]
  {https://doi.org/10.1038/s41592-019-0686-2}, \href {https://rdcu.be/b08Wh}
  {17, 261}

\bibitem[\protect\citeauthoryear{{Vlahakis}, {Dunne}  \& {Eales}}{{Vlahakis}
  et~al.}{2005}]{Vlahakis2005}
{Vlahakis} C.,  {Dunne} L.,   {Eales} S.,  2005, \mn@doi [\mnras]
  {10.1111/j.1365-2966.2005.09666.x}, \href
  {https://ui.adsabs.harvard.edu/abs/2005MNRAS.364.1253V} {364, 1253}

\bibitem[\protect\citeauthoryear{{Vogelsberger}, {Genel}, {Sijacki}, {Torrey},
  {Springel}  \& {Hernquist}}{{Vogelsberger} et~al.}{2013}]{Vogelsberger2013}
{Vogelsberger} M.,  {Genel} S.,  {Sijacki} D.,  {Torrey} P.,  {Springel} V.,
  {Hernquist} L.,  2013, \mn@doi [\mnras] {10.1093/mnras/stt1789}, \href
  {https://ui.adsabs.harvard.edu/abs/2013MNRAS.436.3031V} {436, 3031}

\bibitem[\protect\citeauthoryear{{Vogelsberger} et~al.,}{{Vogelsberger}
  et~al.}{2014a}]{Vogelsberger2014b}
{Vogelsberger} M.,  et~al., 2014a, \mn@doi [\mnras] {10.1093/mnras/stu1536},
  \href {https://ui.adsabs.harvard.edu/abs/2014MNRAS.444.1518V} {444, 1518}

\bibitem[\protect\citeauthoryear{{Vogelsberger} et~al.,}{{Vogelsberger}
  et~al.}{2014b}]{Vogelsberger2014a}
{Vogelsberger} M.,  et~al., 2014b, \mn@doi [\nat] {10.1038/nature13316}, \href
  {https://ui.adsabs.harvard.edu/abs/2014Natur.509..177V} {509, 177}

\bibitem[\protect\citeauthoryear{{Vogelsberger} et~al.,}{{Vogelsberger}
  et~al.}{2020}]{Vogelsberger2020}
{Vogelsberger} M.,  et~al., 2020, \mn@doi [\mnras] {10.1093/mnras/staa137},
  \href {https://ui.adsabs.harvard.edu/abs/2020MNRAS.492.5167V} {492, 5167}

\bibitem[\protect\citeauthoryear{{Weinberger} et~al.,}{{Weinberger}
  et~al.}{2017}]{Weinberger2017}
{Weinberger} R.,  et~al., 2017, \mn@doi [\mnras] {10.1093/mnras/stw2944}, \href
  {https://ui.adsabs.harvard.edu/abs/2017MNRAS.465.3291W} {465, 3291}

\bibitem[\protect\citeauthoryear{{Weinberger} et~al.,}{{Weinberger}
  et~al.}{2018}]{Weinberger2018}
{Weinberger} R.,  et~al., 2018, \mn@doi [\mnras] {10.1093/mnras/sty1733}, \href
  {https://ui.adsabs.harvard.edu/abs/2018MNRAS.479.4056W} {479, 4056}

\bibitem[\protect\citeauthoryear{{Whitaker}, {van Dokkum}, {Brammer}  \&
  {Franx}}{{Whitaker} et~al.}{2012}]{Whitaker2012}
{Whitaker} K.~E.,  {van Dokkum} P.~G.,  {Brammer} G.,   {Franx} M.,  2012,
  \mn@doi [\apj] {10.1088/2041-8205/754/2/L29}, \href
  {https://ui.adsabs.harvard.edu/abs/2012ApJ...754L..29W} {754, L29}

\bibitem[\protect\citeauthoryear{{Wiersma}, {Schaye}, {Theuns}, {Dalla Vecchia}
   \& {Tornatore}}{{Wiersma} et~al.}{2009}]{Wiersma2009}
{Wiersma} R. P.~C.,  {Schaye} J.,  {Theuns} T.,  {Dalla Vecchia} C.,
  {Tornatore} L.,  2009, \mn@doi [\mnras] {10.1111/j.1365-2966.2009.15331.x},
  \href {https://ui.adsabs.harvard.edu/abs/2009MNRAS.399..574W} {399, 574}

\bibitem[\protect\citeauthoryear{{Wiseman}, {Schady}, {Bolmer}, {Kr{\"u}hler},
  {Yates}, {Greiner}  \& {Fynbo}}{{Wiseman} et~al.}{2017}]{Wiseman2017}
{Wiseman} P.,  {Schady} P.,  {Bolmer} J.,  {Kr{\"u}hler} T.,  {Yates} R.~M.,
  {Greiner} J.,   {Fynbo} J.~P.~U.,  2017, \mn@doi [\aap]
  {10.1051/0004-6361/201629228}, \href
  {https://ui.adsabs.harvard.edu/abs/2017A&A...599A..24W} {599, A24}

\bibitem[\protect\citeauthoryear{{Yajima}, {Nagamine}, {Thompson}  \&
  {Choi}}{{Yajima} et~al.}{2014}]{Yajima2014}
{Yajima} H.,  {Nagamine} K.,  {Thompson} R.,   {Choi} J.-H.,  2014, \mn@doi
  [\mnras] {10.1093/mnras/stu169}, \href
  {https://ui.adsabs.harvard.edu/abs/2014MNRAS.439.3073Y} {439, 3073}

\bibitem[\protect\citeauthoryear{{Zafar} \& {Watson}}{{Zafar} \&
  {Watson}}{2013}]{Zafar2013}
{Zafar} T.,  {Watson} D.,  2013, \mn@doi [\aap] {10.1051/0004-6361/201321413},
  \href {https://ui.adsabs.harvard.edu/abs/2013A&A...560A..26Z} {560, A26}

\bibitem[\protect\citeauthoryear{{Zhukovska}, {Dobbs}, {Jenkins}  \&
  {Klessen}}{{Zhukovska} et~al.}{2016}]{Zhukovska2016}
{Zhukovska} S.,  {Dobbs} C.,  {Jenkins} E.~B.,   {Klessen} R.~S.,  2016,
  \mn@doi [\apj] {10.3847/0004-637X/831/2/147}, \href
  {https://ui.adsabs.harvard.edu/abs/2016ApJ...831..147Z} {831, 147}

\bibitem[\protect\citeauthoryear{{Zinger} et~al.,}{{Zinger}
  et~al.}{2020}]{Zinger2020}
{Zinger} E.,  et~al., 2020, arXiv e-prints, \href
  {https://ui.adsabs.harvard.edu/abs/2020arXiv200406132Z} {p. arXiv:2004.06132}

\bibitem[\protect\citeauthoryear{{da Cunha}, {Charlot}  \& {Elbaz}}{{da Cunha}
  et~al.}{2008}]{daCunha2008}
{da Cunha} E.,  {Charlot} S.,   {Elbaz} D.,  2008, \mn@doi [\mnras]
  {10.1111/j.1365-2966.2008.13535.x}, \href
  {https://ui.adsabs.harvard.edu/\#abs/2008MNRAS.388.1595D} {388, 1595}

\bibitem[\protect\citeauthoryear{{da Cunha} et~al.,}{{da Cunha}
  et~al.}{2013}]{daCunha2013}
{da Cunha} E.,  et~al., 2013, \mn@doi [\apj] {10.1088/0004-637X/766/1/13},
  \href {https://ui.adsabs.harvard.edu/abs/2013ApJ...766...13D} {766, 13}

\bibitem[\protect\citeauthoryear{{van der Walt}, {Colbert}  \&
  {Varoquaux}}{{van der Walt} et~al.}{2011}]{Numpy}
{van der Walt} S.,  {Colbert} S.~C.,   {Varoquaux} G.,  2011, Computing in
  Science Engineering, 13, 22

\makeatother
\end{thebibliography}



\appendix

\section{Stellar mass distributions}\label{app:A_Ms}

\begin{figure*}
	\includegraphics[width=0.98\textwidth]{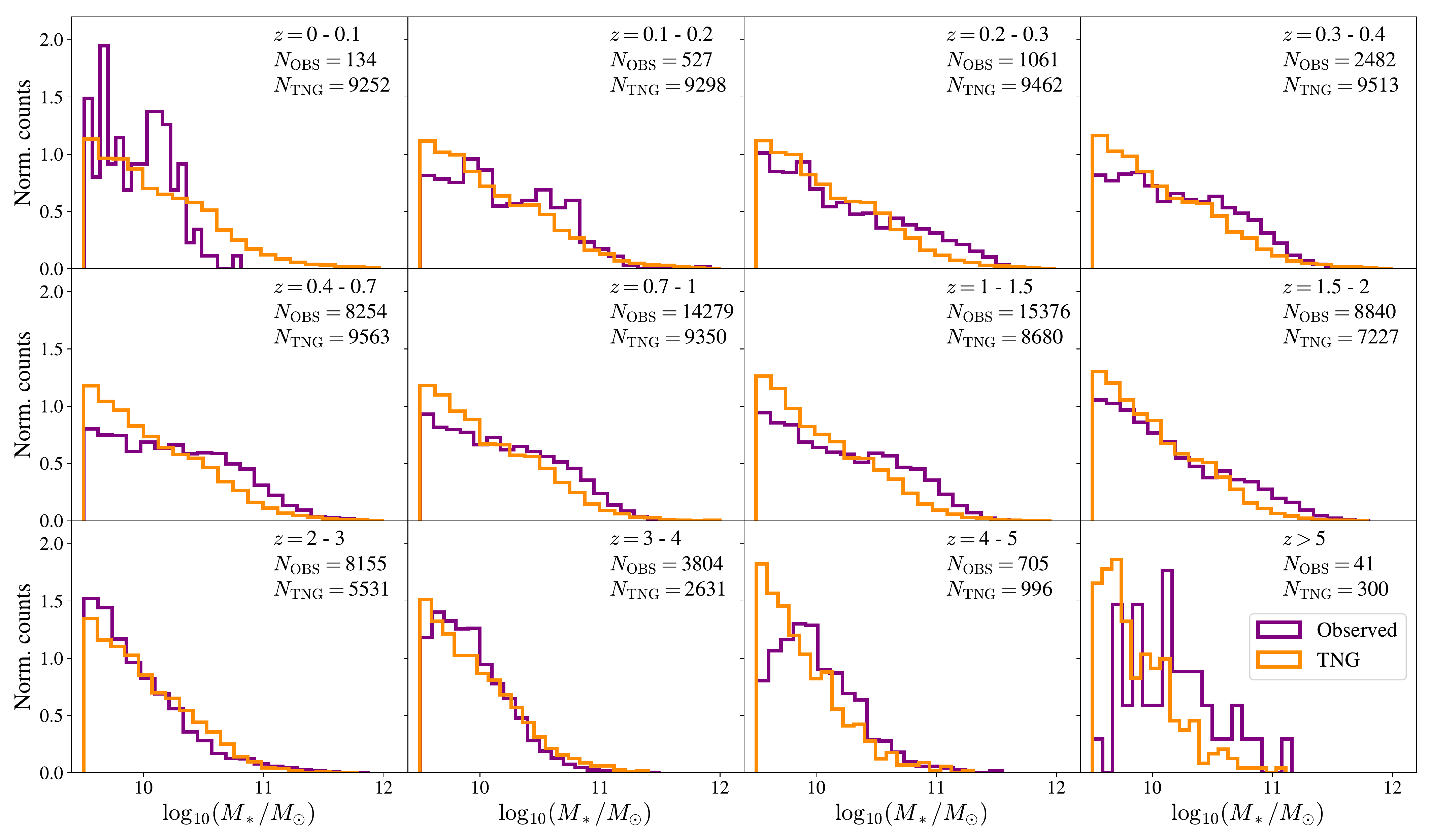}
    \caption{Distributions of the stellar masses of galaxies in different redshift bins for TNG100 and {\sc magphys}-COSMOS sources. TNG100 stellar masses are calculated using particles and cells within $2r_{0.5}$. \textit{Purple:} {\sc magphys}-COSMOS sources used in stacking analysis to determine average dust properties. \textit{Orange:} TNG100 galaxies.}
    \label{fig:mstar_hists_stacking_TNG}
\end{figure*} 

\section{TNG100 dust-to-stellar-mass ratios}\label{app:B_TNG_MdMs}

\begin{figure*}
	\includegraphics[width=0.98\textwidth]{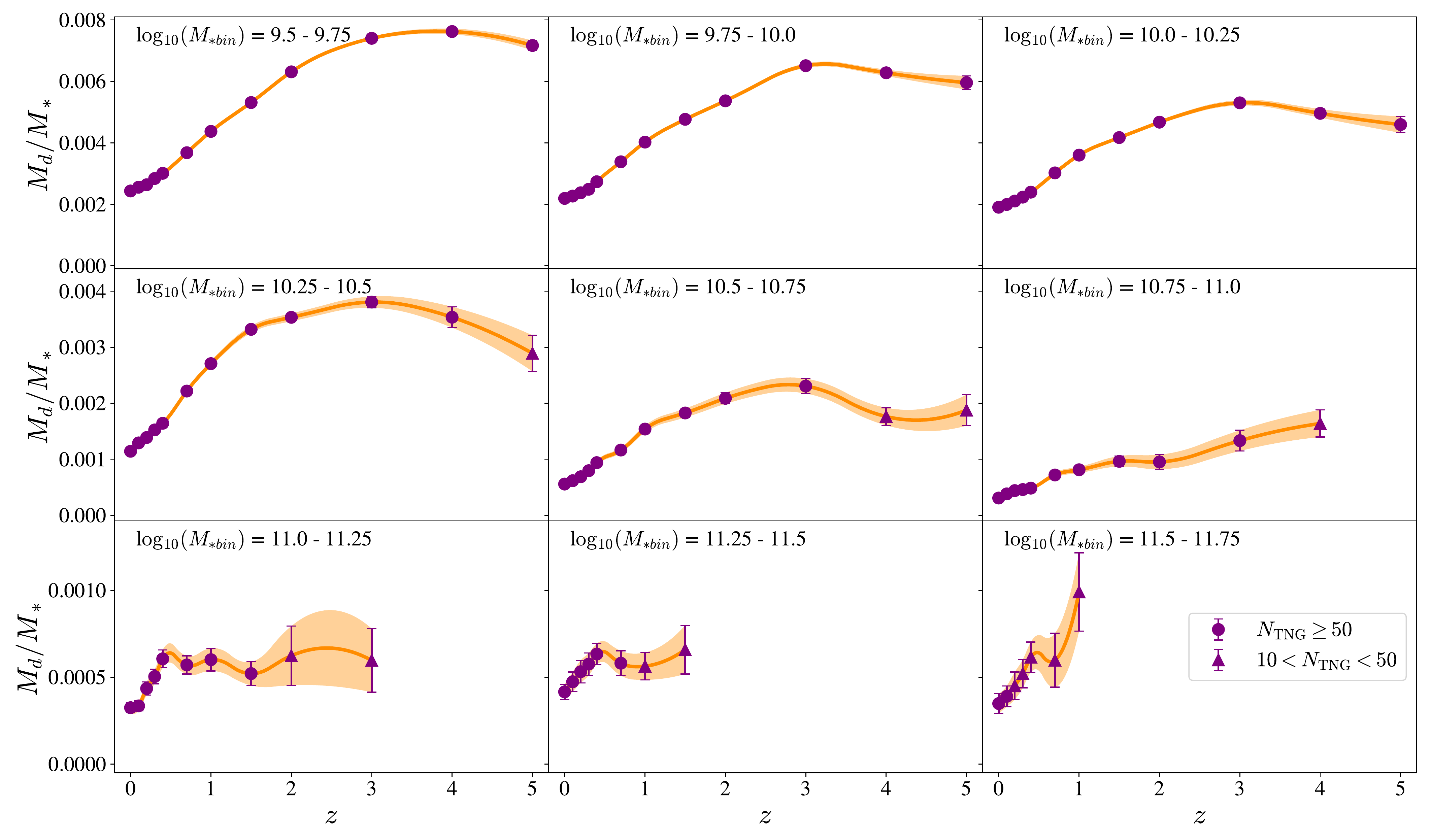}
    \caption{Dust-to-stellar-mass ratios obtained from TNG100 simulations in redshift and stellar mass bins. \textit{Circles:} at least 50 sources in the $(M_* - z)$ bin. \textit{Triangles:} between 10 and 50 sources in the $(M_* - z)$ bin. The line is a quadratic interpolation to the data. Note that the scale on the y-axis has been chosen to make clear the evolution that is predicted by TNG100 and is not the same as the scale used in Figure \ref{fig:stacked_TNG_evolution}.}
    \label{fig:TNG_evolution_interp}
\end{figure*} 


\bsp	
\label{lastpage}
\end{document}